\documentclass[aps,prb,twocolumn,superscriptaddress,showpacs,english,citeautoscript]{revtex4-1}

\usepackage[T1]{fontenc}
\usepackage[latin9]{inputenc}
\usepackage{graphicx}
\usepackage{multirow}
\usepackage[version=3]{mhchem}
\usepackage[dvipsnames]{xcolor}
\usepackage[linktocpage=true,
  colorlinks=true, 
  pdfborder={0 0 0},
  linkcolor=blue,
  citecolor=blue,
  filecolor=yellow,
  urlcolor=blue,
  bookmarks,
  pdfauthor={Flores},
]{hyperref}

\newcommand{\mapbi}{MAPI }
\newcommand{\ma}{methylammonium }
\newcommand{\BaselPhys}{Department of Physics, Universit\"at Basel, Klingelbergstr. 82, 4056 Basel, Switzerland} 
 
\newcommand{\EPFL}{Laboratory of Computational Chemistry and Biochemistry, 
Institute of Chemical Sciences and Engineering, Ecole Polytechnique F\'ederale de Lausanne, CH-1015 Lausanne, Switzerland}
\newcommand{\LASSP}{Laboratory of Atomic and Solid State Physics, Cornell University, Ithaca, New York 14853, USA}

\begin{document}

\title{Emergence of hidden phases of methylammonium lead-iodide (CH$_3$NH$_3$PbI$_3$) upon compression}

\author{Jos\'e~A. Flores-Livas}  \affiliation{\BaselPhys}\email{jose.flores@unibas.ch}
      \author{Daniele Tomerini}  \affiliation{\BaselPhys}
     \author{Maximilian Amsler}  \affiliation{\LASSP}
        \author{Ariadni Boziki}  \affiliation{\EPFL}
 \author{Ursula Rothlisberger}   \affiliation{\EPFL}
      \author{Stefan Goedecker}  \affiliation{\BaselPhys}\email{Stefan.Goedecker@unibas.ch}

\date{\today}

\begin{abstract}
We perform a thorough structural search with the minima hopping method (MHM) to explore low-energy structures of \ma lead iodide. 
By combining the MHM with a forcefield, we efficiently screen vast portions of the configurational 
space with large simulation cells containing up to 96 atoms.  
Our search reveals two structures of \ma iodide perovskite (MAPI) that are substantially lower in energy 
than the well-studied experimentally observed low-temperature $Pnma$ orthorhombic 
phase according to density functional calculations. 
Both structures have not yet been reported in the literature for MAPI, but our results show that they 
could emerge as thermodynamically stable phases via compression at low temperatures.  
In terms of the electronic properties, the two phases exhibit larger band gaps than the standard perovskite-type structures. 
Hence, pressure induced phase selection at technologically achievable pressures (i.e., via thin-film strain) 
is a route towards the synthesis of several \mapbi polymorph with variable band gaps.
\end{abstract}

\maketitle

\section{Introduction}

Perovskites are among the most promising and versatile class of candidate compounds for new or improved materials in energy applications, 
including photovoltaics, superconductivity and lasing~\cite{alivisatos_god-perovska2017}. 
With the general formula \ce{ABX3}, the perovskite structure consists of corner-sharing \ce{BX6} octahedra forming a 3D framework
that provides room for the A units in the resulting cuboctahedral cavities. 
Materials adopting the perovskite structure can display many desirable properties,  
that are unparalleled among all other families of compounds~\cite{10.1038/nature00893,doi:10.1021/jz5005285,
PhysRevB.63.113104,10.1038/nature12509,Lee02112012,10.1038/nature05023,10.1038/nature02572,cmr,1987ApPhL..51...57B,
PhysRevLett.58.908,Tateyama_termination_2014,report-piezoelectrics}.  
This wide range of properties arises from the large number of elements that can be accommodated in the
rather {\it simple} crystal structure. Further, the high-symmetry cubic perovskite structure 
is flexible and can readily distort, either as an effect of temperature and pressure, or of intercalation, doping and defects~\cite{mitchell2002perovskites}, leading to orthorhombic, tetragonal, or trigonal type perovskites. 
 
The lead-based \ma iodide perovskite (MAPI), with A = \ce{CH3NH3+},
has a band gap of $\sim$\,1.55\,eV at room temperature, which 
is close to the optimal range of 1.1 to 1.4\,eV for a single junction photovoltaic cell~\cite{kojima_JACS}. 
Since its introduction as sensitizer for liquid-electrolyte based dye-sensitized solar cells in 2009~\cite{kojima_JACS}, 
\mapbi has emerged as one of the most cost effective materials for photovoltaic applications, with continuously and rapidly improving efficiency. 
Rapid progress has been achieved within an extremely short period of time, and is a direct result of intensive studies~\cite{oba2018design} 
involving improved designs of the photovoltaic cell architecture, optimization of the preparation procedure, 
as well as modifications of optical and electronic characteristics through chemical substitution. 
Meanwhile, the power conversion efficiency of \mapbi has almost reached  20\,\%, closer and closer to the Shockley-Queisser limit. 

One of the most technologically relevant features of \mapbi is its high optical absorption while incorporated in typical 
hybrid perovskite solar devices. In many of these devices, the \mapbi absorbers are grown as homogeneous, 
ultra-thin films with a thickness of up to a few hundred nanometers, sandwiched between conductive layers. 
The strain arising at the interface between the \mapbi and the substrate can distort the perovskite structure,
and affect the opto-electronic properties. This has been shown for superlattices of BaTiO$_3$, 
for which the interface structure enhances the polarization~\cite{lee2005strong}. 
In turn, a careful selection of lattice mismatch to induce compressive and tensile strains has been proposed 
as an alternative route to tune the electronic properties of perovskite materials 
without affecting the chemical composition. 

\begin{table}[htb]
\caption{~Effect of different exchange-correlation functionals on the lattice parameters and the volume of the orthorhombic 
($Pnma$) structure at zero pressure. Experimental values stem from reference~\onlinecite{whitfield_NeutronDif_2016}.}
\begin{tabular} {l|ccccr} 
Functional &   $a$   &   $b$   &   $c$   & Volume   &   T  \\ [-2.pt] 
           &   (\AA) &   (\AA) &   (\AA) &(\AA$^3$) &  (K)  \\ \hline\\[-8.pt]
Exp.     &  8.81 & 8.55 & 12.58 & 949  &  10 \\ 
Exp.     &  8.86 & 8.57 & 12.62 & 960  & 100 \\ 
Exp.     &  8.86 & 8.58 & 12.62 & 960  & 150 \\ \hline
LDA      &  8.71 & 8.30 & 12.43 & 900  & -  \\
PBE      &  9.24 & 8.63 & 12.91 & 1031 & -  \\
PBEsol   &  8.99 & 8.45 & 12.61 & 959  & -  \\ \hline 
DFT-D2   &  8.68 & 8.57 & 12.60 & 939  & -  \\ 
DFT-D3   &  8.91 & 8.54 & 12.75 & 971  & -  \\
T-S      &  8.97 & 8.49 & 12.73 & 972  & -  \\
MB-d     &  8.97 & 8.51 & 12.73 & 973  & -  \\ 
dDsC     &  8.96 & 8.52 & 12.71 & 972  & -  \\ \hline
PBE0     &  9.03 & 8.65 & 12.93 & 1012 & -  \\ 
SCAN     &  8.95 & 8.62 & 12.73 & 983  & -  \\  
\end{tabular} \label{table:fxc_orthorombic}
\end{table}

The pressure-temperature phase diagram of \mapbi is however rather complex. 
At ambient pressure and low temperature, \mapbi adopts an orthorhombic structure with $Pnma$ symmetry~\cite{whitfield_NeutronDif_2016}, 
which can be assigned the  $a^- b^+ a^-$ label in Glazer notation~\cite{glazer_notation_1972}. 
At 162\,K, the $Pnma$ phase transforms via a first order phase transition involving a rearrangement of the octahedra 
to a tetragonal cell with $I4/mcm$ symmetry, $a^0 a^0 c^-$, which persists at ambient temperature. 
Upon heating above 327\,K~\cite{whitfield_NeutronDif_2016,capitani2016high}, \mapbi transforms to a cubic structure with 
$Pm\text{--}3m$ symmetry with the $a^0 a^0 c^0$ classification. 

Exploring the phase diagram of hybrid organic-inorganic perovskites under pressures is challenging, and available data points towards 
a complex enthalpy landscape rich in crystallographic transformations. Recently, state-of-the-art synchrotron powder X-ray diffraction 
experiments on \mapbi under pressure were reported, however without conclusively resolving the complete crystal structure~\cite{capitani2016high}. 

Some of the experimental results disagree with each other on the nature of transformations occurring at moderate pressures. 
Szafra\'nki~\textit{et al.}~\cite{szafranski2016mechanism,szafranski2017photovoltaic} 
on one hand suggested that \mapbi undergoes amorphization above 2.5\,GPa. 
On the other hand, a transformation to a post-tetragonal phase was observed by both 
Capitani and Jiang~\textit{et al.}~\cite{capitani2016high,Jiang_pressuremapbi_angew}. 
A third set of experiments suggests that mixed organic-inorganic perovskites remain stable in their ambient 
(or a distorted) form and metalize for pressures above 50\,GPa~\cite{jaffe2016high,jaffe2017halide}.  
These discrepancies clearly show that further investigations are called for to conclusively map out 
the high-pressure phase diagram of MAPI.

\begin{figure*}[htb] 
\includegraphics[width=1.0\linewidth,angle=0]{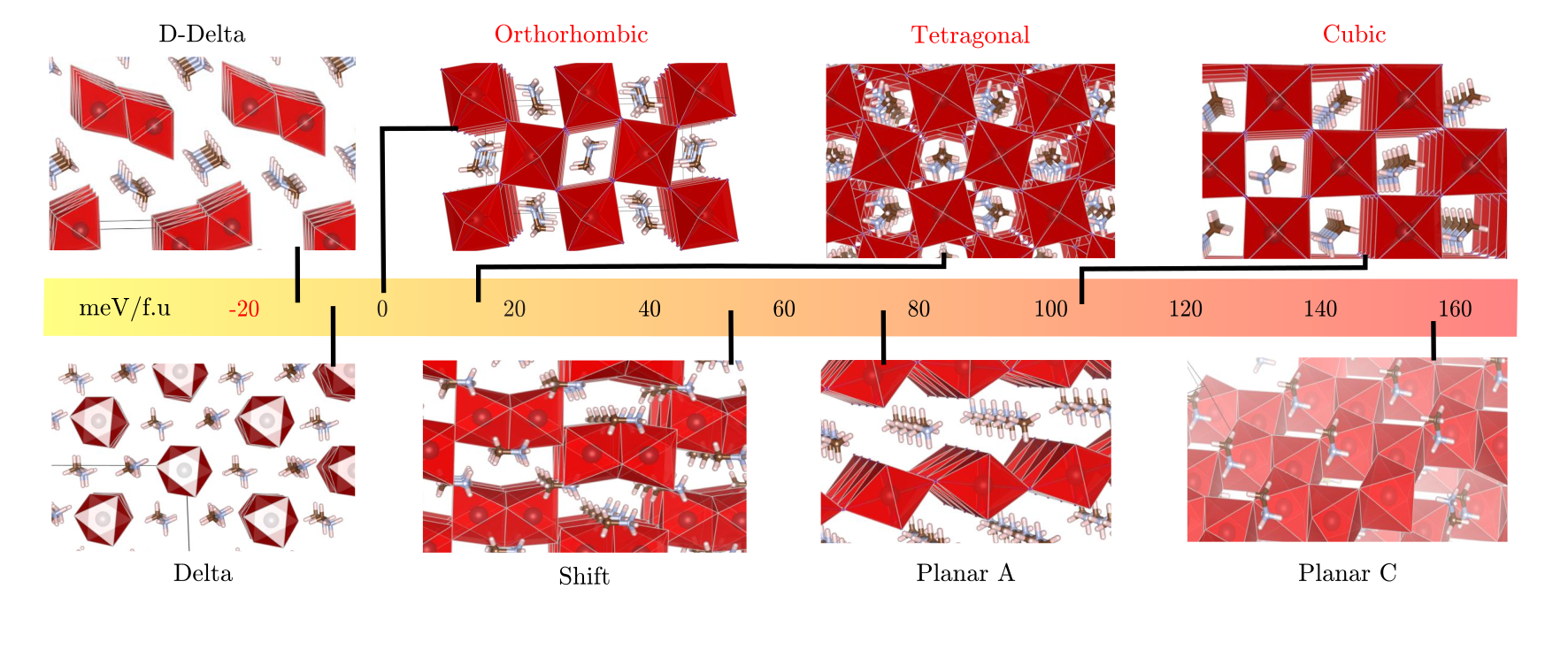} 
\caption{Energy ranking of various \mapbi polymorphs with respect to the orthorhombic ($Pnma$) 
phase with DFT using the meta-GGA (SCAN + SOC) functional. 
Structural motifs include the known perovskites (cubic, tetragonal and orthorhombic) 
as well as the low-lying polymorphs predicted in this work.} 
\label{fig:structures}
\end{figure*}

One way to tackle this challenge is to use computational modeling based on first principles calculations 
in conjunction with structural search algorithms to explore potential high-pressure phases. 
However, the modeling of \mapbi using density functional theory (DFT) is challenging on its own. 
Standard exchange-correlation functionals used in DFT, such as the generalized gradient approximation, 
are known to underestimate the band gaps of MAPI, and structural features like the equilibrium volumes deviate 
from experimental values by up to 10\,\%. 
This discrepancy can be reduced by using improved functionals, e.g., by including empirical van der Waals (vdW) interactions. 
Further, spin-orbit effects are known to play a role in accurately reproducing experimental band gaps~\cite{even2013importance}. 
Despite these challenges, there have been several attempts to explore the chemical and structural landscape of 
perovskite materials, e.g., using approaches based on high-throughput schemes~\cite{korbel2018stable,emery2016high},  
and machine learning~\cite{schmidt2017predicting,sarmiento2015prediction}. 

The most unconstrained manner to explore the potential energy surface (PES) is to employ a sophisticated structural search 
algorithm that samples the configurational space at a given stoichiometry. To the best of our knowledge, 
only Huan~\textit{et al.}~\cite{Huan_PhysRevB,Kim_SciData} used this approach, by employing the minima hopping method 
(MHM)~\cite{MHM,Amsler_2010} to directly explore the DFT potential energy surface (PES) of several halide perovskites 
(in particular, 12 atoms, or 1 f.u. for MAPI). This study reports interesting unobserved structures that are 
structurally related to the experimentally known, cubic high-temperature $Pm\text{--}3m$ phase. 
To correctly model the complex low-temperature orthorhombic phase (which has a significantly lower energy 
than the structures predicted by Huan~\textit{et al.}), a system size of at least 4 f.u. per cell (48 atoms per cell) is required 
to allow for different tilting angles of the octahedra. 
Since the complexity of the energy landscape grows exponentially with the number of atoms, 
a structural search directly at the DFT level with such a large cell is computationally prohibitive. 

In this work, we address this issue by screening the PES of \mapbi using a classical force-field to model large supercells 
in conjunction with the MHM. We identify two unreported polymorphs of \mapbi that are considerable lower in energy than any 
of the perovskite (cubic, tetragonal and orthorhombic) phases. 
Further both structures are clearly thermodynamically favored at increased pressures, 
and exhibit systematically larger band gaps than the know perovskites. 
The manuscript is organized as follows. Section~\ref{sec:results} consists of three subsections: 
the first section~\ref{sec:novel} summarizes the structural search and discusses the thermodynamic stability of the low-energy phases. 
Section~\ref{sec:pressure} investigates the effect of pressure and stability upon compression, 
while section~\ref{sec:electronic} assesses the electronic properties. 
Finally, section~\ref{sec:discussion} discusses potential routes towards the synthesis of the new phases of MAPI. 

\section{Results\label{sec:results}}

\subsection{Novel phases of \mapbi\label{sec:novel}}

We use a classical inter-atomic potential parametrized by Handley and Freeman~\cite{handley_gulppotential}, as implemented in the General Utility Lattice Program (GULP) molecular modeling package~\cite{curtin_gulpref}. 
GULP was coupled to the MHM to screen the PES using simulation cells containing 2, 4 and 8  units (i.e., 24, 48 and 96 atoms). 
More than 140,000 polymorphs of \mapbi are generated using this procedure. 
To validate the accuracy of the force-field, we randomly select 250 structures and optimize them using KS-DFT with (LDA and PBE) functionals.
We find a remarkably good correlation between the DFT and force-field energies 
(see Supplemental Information, SI), which justifies its use for a reliable prediction of low-energy structures. 

To assess which exchange-correlation functional is best suited to model \mapbi, 
we compute the lattice parameters and cell volumes of the experimental ground state, the orthorhombic phase with 48 atoms, 
using a set of different functionals (see  Table~\ref{table:fxc_orthorombic}). We employ three popular standard functionals:  
the local-density approximation (LDA)~\cite{Perdew-Wang_1992} 
and two generalized gradient approximation (GGA) (PBE~\cite{PBE96} 
and PBEsol~\cite{PBESol}). 
We also assess functionals that (empirically) take into account vdW dispersion, 
namely the DFT-D2 method~\cite{grimme2006_D2} and the zero damping DFT-D3 method of Grimme~\cite{grimme2010_D3}, 
the Tkatchenko-Scheffler (T-S) method~\cite{tkatchenko_first_2009}, and 
the many-body dispersion (MB-d) energy method~\cite{manyBody_cor_tkatchenko2012}, 
and the dispersion correction (dDsC) method~\cite{dDsC_vdW_2011}. 
We also investigate a hybrid functional which includes a fraction of exact-exchange, PBE0~\cite{PBE0}, 
and the recently developed strongly constrained and appropriately normed (SCAN) meta-GGA functional~\cite{sun2016accurate}.

As expected, the standard Kohn-Sham functionals provide a fair description of the lattice constants 
when compared to experiments~\cite{Rinke_eggsVDW_2016,benchmark_pnma_2014}. 
LDA and PBE under- and overestimate the lattice constants, respectively, and PBEsol, while at first glance 
provides cell volumes in excellent agreement with experiment, exhibits a deviation of the individual lattice components. 
Empirical inclusion of vdW interactions greatly improves the unit cell volumes (D2 and D3)~\cite{mamon-Giustino_anaharmonic_2017}). 
The D2 and D3 approximations show $b$-components in excellent agreement with experimental values~\cite{wiktor2017predictive}, 
display similar trends. All vdW methods lead to very accurate $c$-component when compare to experiments. 
Surprisingly, the inclusion of exact-exchange via hybrid-DFT leads to a systematic overestimation 
of the three lattice parameters, while being the computationally most costly method studied here. 
Finally, our analysis indicates that the meta-GGA SCAN functional improves the overall description 
with respect to experimental values compared to other semi-local based functionals. 
It is worth mentioning that the volume obtained by SCAN is reliably close to the experimental 
ones~\cite{scan_benchmark_2018,MD_Jonathan_PRB-2016,PRL_Menno-Jonathan_RPA-SCAN_2017} even without the inclusion of 
vdW correction (and will become relevant under pressure), and represents a good trade-off between accuracy and computational cost. 
Therefore, henceforth we will use the SCAN functional for all structural and thermodynamic evaluations.

First, we compute the thermodynamic stability of the various polymorphic modifications of 
\mapbi based on the SCAN functional. We identify several distinct structural motifs within a small energy range around 
the ground state phase which we consider in our analysis. Fig.~\ref{fig:structures} shows representative structures of various low-energy phases, 
ranked according to their energies with respect to the orthorhombic structure of MAPI. 
We use our recently developed structural fingerprint~\cite{zhu2016fingerprint,flores2017accelerated} 
to classify the large amount of distinct phases into families of similar structural motifs (see SI).
One family that we call "planar" encompasses all structures with layered octahedra that share edges and are 
intercalated with molecular units of \ce{CH3NH3+} between these planes. 
The structure referred to as "planar A" lies between the tetragonal and the cubic phase with an energy of $\sim$\,75\,meV/f.u. 
above the ground state, while "planar C" lies at an energy scale of $\sim$\,125\,meV/f.u. ("planar B" is not shown in the figure). 
Considering the energy range spanned by the experimentally observed cubic and orthorhombic phases of $\sim$\,105\,meV/f.u.,  
"planar C" is unlikely to be observed in experiments, while "planar A" lies readily within the range of the 
thermodynamic scale of metastable phases and is hence potentially 
accessible~\cite{Sun2016thermodynamic,amsler2018exploring}. 

Another structural motif that we find and define as the "shift" family consists of \ce{PbI6} octahedra arranged in $(ch)_4$ fashion 
according to the $h-c$ notation~\cite{Jagodzinski:a00148,tilley2016perovskites}. 
These phases have energy difference with the ground state that are at least twice that of the tetragonal phase, but lower than the cubic phases. 
The organic components are arranged in a way as to compensate for the change in the back-bone structure of the perovskite type phase.
Hence, this shift type of structural motif could be energetically further enhanced if other types of cations are selectively used. 
Other arrangements of hexagonal and cubic layers such as $(cchh)_2$ were found in our forcefield search, but not included in further analysis 
as they were found to be energetically higher.

Finally, and most significantly, we report here the discovery of two classes of 
polymorphs which lie -15 and -9~meV/f.u. lower in energy than the reference orthorhombic phase, 
namely the double-delta (D-Delta) and delta structures. 
The unit cell of the delta structure contains 48 atoms, and consists of face-sharing octahedra stacked on 
top of each other to form pillars, while the organic parts uniformly surround them. 
In this structure, the Pb-I distance within the octahedral arrangements is on average 3.25\,\AA, 
and the distance between the pillars and the molecules are on average 3.6\,\AA. 
The D-delta phase also has a pillar-like structure, but the octahedra share 
edges instead of faces. We selected three structures (that we labeled --A: 48 atoms and --B:, --C: 96 atoms) 
with different orientation of the \ma molecules; the difference are shown in the supplementary information. 
The average Pb-I distance in D-delta is 3.27\,\AA, and the distance from "double-pillars" 
to the molecules is on average 3.7\,\AA. The coordination number per octahedra 
(i.e., the number of iodide ions linked to Pb centers of the octahedra) is nominally close to 6 in both structures (delta and D-delta). 
Note that in all the structures described here, all molecular \ce{CH3NH3+} remain intact: the inter-atomic distances and the overall 
conformation are preserved. The main structural differences between the structures arise from the different octahedral 
conformations and the rearrangement of molecular cations.

\subsection{\mapbi under pressure\label{sec:pressure}}
\begin{figure}[t!]
\includegraphics[width=1.0\linewidth,angle=0]{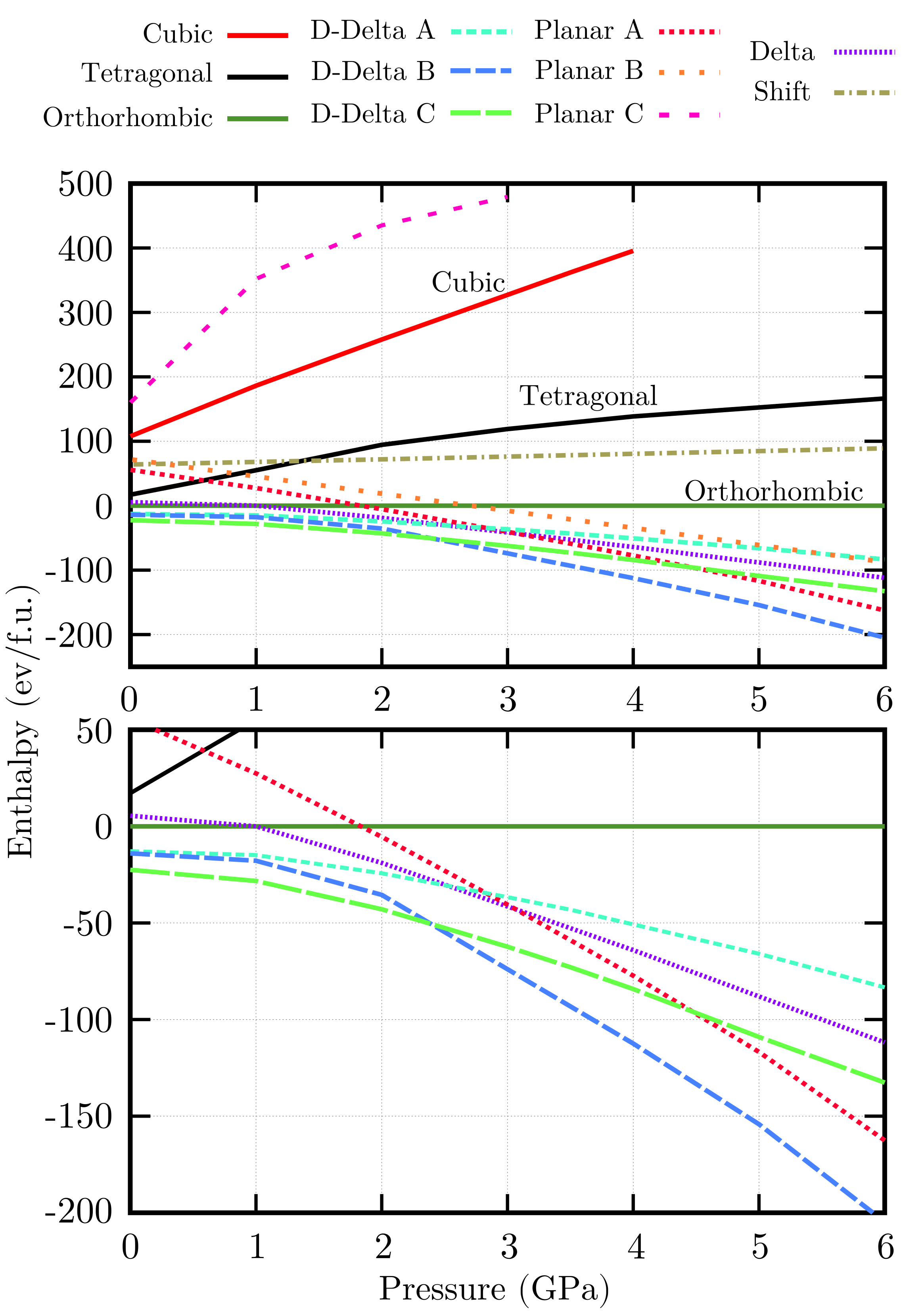}
\caption{~Calculated formation enthalpies per formula unit (12 atoms) of 
various competing \mapbi polymorphs as a function of pressure, computed using the meta-GGA functional (SCAN + SOC). 
Bottom panel is a magnification of the low-enthalpy region. All enthalpy differences are given with respect to the orthorhombic ($Pnma$) phase.}
\label{fig:PV}
\end{figure}

We next investigate how pressure affects the enthalpy and volume of the experimental and  predicted phases of \mapbi with respect to the orthorhombic phase effect, and we report the results in Fig.~\ref{fig:PV}. Clearly, the tetragonal and cubic phases are greatly destabilized upon compression, 
while other phases such as planar A and B and the delta phases become enthalpically preferable over the orthorhombic phase. 
Within the range of pressure that we study here we find that the delta phases, in particular the D-Delta structures, 
are predominantly stabilized upon compression. 
More specifically, we note that several D-Delta phases, which all contain distinctly oriented organic cations, leave the inorganic framework completely intact upon compression. The D-Delta C phase is thermodynamically stable between zero and 2.5\,GPa, above which the D-Delta B structure is favored. 
This phase exhibits a more symmetric arrangement of the organic moieties, and persists as the thermodynamic ground state up to at least 6\,GPa. 
Both the B and C phases are members of the double-delta family, which appears as the most stable class of structure at all pressures.
Therefore it appears that the face- and edge-sharing arrangement of the octahedra, being more compact, is favored over a less compact, 
corner-sharing arrangement. The enthalpy differences between the distinct double-delta phases (A, B, and C) primarily stem from different orientations 
of the molecular parts in the space between the octahedral frameworks, which interact primarily through different dipole arrangements. 
A clear enthalpic preference simply based on the relative unit cell volumes is therefore not possible.

\begin{figure}[t!]
\includegraphics[width=1.05\linewidth,angle=0]{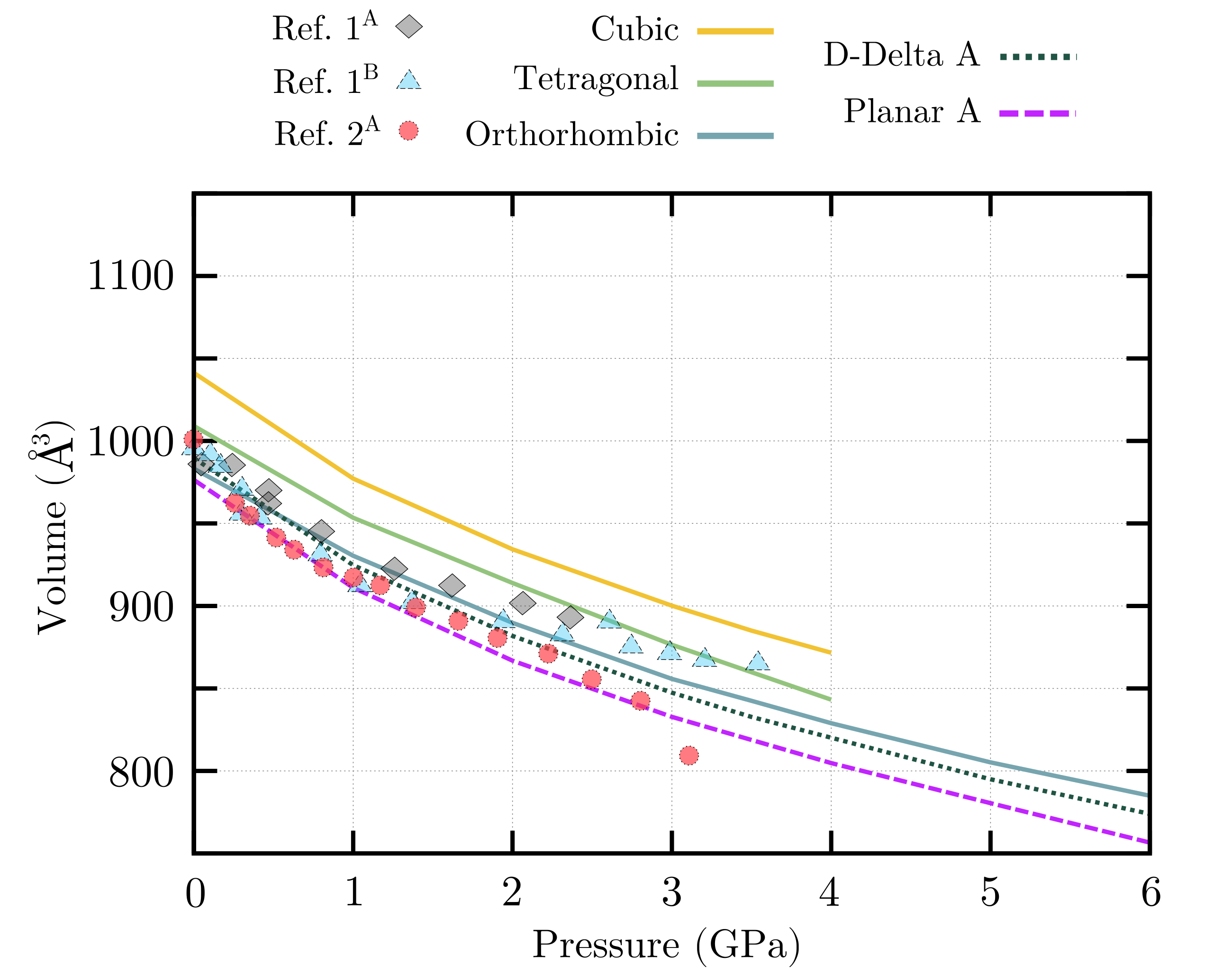}
\caption{~Pressure-volume dependence for various predicted polymorphs of \mapbi
and experimental values from Ref.~1 (\onlinecite{Jiang_pressuremapbi_angew}) and Ref.~2 (\onlinecite{capitani2016high}).}
\label{fig:volume}
\end{figure}

In Fig.~\ref{fig:volume}, we map out the pressure dependence of the atomic volumes of the various predicted polymorphs of \mapbi and compare them with experimentally available data. The cubic phase has the largest volume of all experimentally observed perovskite structure types,  
followed by tetragonal and orthorhombic as more compact phases. 
The room temperature compression carried out by Jiang et al.~\cite{Jiang_pressuremapbi_angew} reports 
the volume for the tetragonal perovskite (see supplemental Material XRD section) and its partial transformation to a $Im$--$3m$ structure 
above 0.4\,GPa. Both phases coexist with each other up to 2.5\,GPa, and at 2.7\,GPa an additional phase appears, whose crystal structure could not be determined. The authors however proposed an indexing with the $Immm$~\cite{swainson2007pressure,wang2015pressure,Jiang_pressuremapbi_angew} space group. Finally, the onset of amorphization occurs at pressures exceeding 4.7\,GPa. 

Capitani et al.~\cite{capitani2016high,postorino2017pressure} also reported  
a similar pressure-volume curve as in one of the experiment Jiang et al. (1B in the figure) , but with a clear deviation 
above 2\,GPa towards smaller volumes. The authors interpreted this effect as 
the emergence of amorphization, a reversible process that leads to a recrystallization of the orthorhombic $Imm2$ upon pressure release. 
However, from the computed volumes of our predicted phases one can see that the volume of planar-A, delta and 
double-delta, are  all smaller than the one of the  orthorhombic (perovskite) phases due to the optimized conformation of octahedral sharing. 
Hence, these phases serve as potential candidate phases that would explain the experimental results of Capitani et al.

\subsection{Electronic structure\label{sec:electronic}}

We characterize the known and predicted phases in terms of their electronic properties under compression. 
In the past, multiple studies have shown that, in order to have a correct description of the electronic properties of \mapbi from first-principles, 
it is important to take into account spin-orbit coupling (SOC)~\cite{umari_GWSOC_2014,wiktor2017predictive,benchmark_pnma_2014}. 
The calculated GGA-PBE scalar-relativistic band gaps of \mapbi polymorph
(tetragonal at experimental volume) is on average close to 1.6\,eV, 
a value that is coincidentally very close to the experimental value of 1.682\,eV~\cite{blancon2017extremely}. 
This is an artifact due to the neglect of SOC of the Pb and I atoms 
in the calculation~\cite{umari_GWSOC_2014,wiktor2017predictive}. 
When SOC is taken into account, the calculated band gap within fully relativistic DFT is $\sim$~0.6\,eV (see Supplemental information). 
In order to address this issue, we use the relaxed geometries calculated with meta-GGA (SCAN) and 
perform single-point calculations with hybrid-DFT (PBE0) including SOC
to accurately estimate the electronic band gaps (see benchmarks and details in Supplemental Materials). 

Fig.~\ref{fig:gap} shows the evolution of the calculated band gaps under pressure for the 
three know polymorphs and selected predicted structures of MAPI. 
Most of the structures have direct transitions (optical band gaps), 
which are shown as continues lines, while those with indirect gaps are shown with dashed lines.  
Experimental photoluminescence measurements of the band-gaps under pressure are included 
for the tetragonal phase~\cite{Jiang_pressuremapbi_angew}.  
The experimental values reported in the literature at ambient pressures for the three perovskite polymorphs are scattered 
between 1.55 to 1.72\,eV
~\cite{dar2016origin,eperon2014formamidinium,baikie2013synthesis,milot2015temperature,motta2015revealing,blancon2017extremely,leguy2016experimental,kim2017direct}. 
We assign this large spread to several potential sources, 
(a) to the range of temperature at which the polymorphs are reported to coexist~\cite{whitfield_NeutronDif_2016}, 
(b) to different orientations of the \ma units (leading to changes of the band gap within a few tens of meV), 
(c) due to errors arising from the method of the measurements (optical absorption edge, photoluminescence, electrical measurements), 
and (d) to the macroscopic conformation of the sample (single crystal of polycrystalline heterojunctions). 
Accordingly to our calculations, the Delta phase has an indirect band gap of 3.81\,eV at zero pressure, which decreases to 3.26~eV at 6\,GPa. 
The double-delta "A" phase has a direct band gap of 3.45\,eV at zero pressure that decreases to 3.12 eV at 6\,GPa. 
The distinct alignment of organic parts for the double-delta phases can modify up to 100\,meV the electronic band gap.

\begin{figure}[t!]
\includegraphics[width=1.0\linewidth,angle=0]{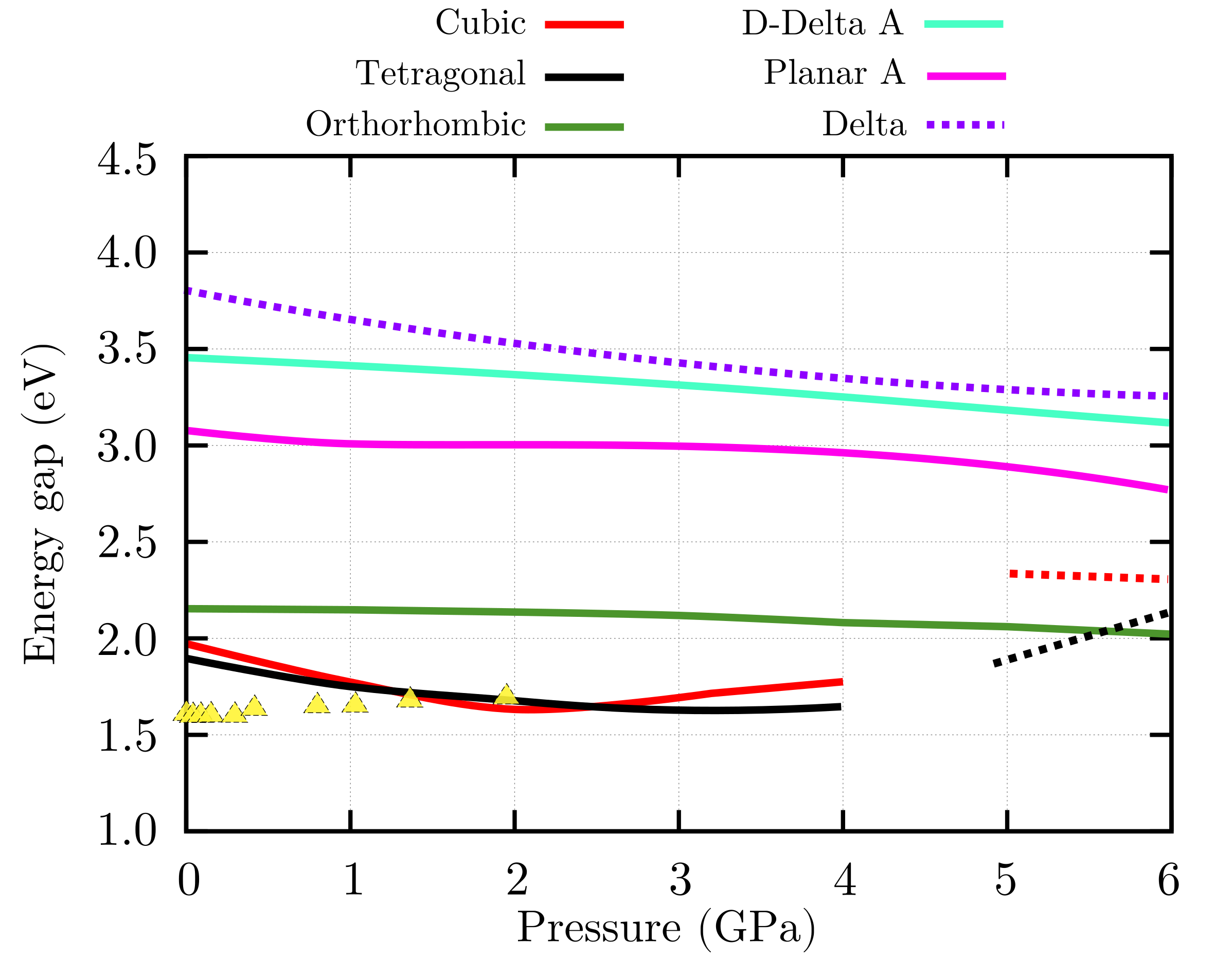}
\caption{~Electronic band gaps calculated with the hybrid functional PBE0, incorporating spin-orbit effects 
(SOC) on optimized geometries from meta-GGA for known and predicted phases of \mapbi under pressure.
Direct band gaps are depicted as solid lines, while dashed lines denote indirect band gaps. 
The yellow dots represent experimental (photoluminescence) values from Ref.~\onlinecite{Jiang_pressuremapbi_angew}.} 
\label{fig:gap}
\end{figure}

\section{Discussion\label{sec:discussion}}

First, let us address in more detail whether the predicted, new phases of \mapbi can be synthesized. 
As previously described, the known structural transitions in hybrid perovskites at ambient pressure are driven by temperature. 
Below 132\,K, the tetragonal phase transforms to the low-temperature orthorhombic phase, 
which persists at least down to 10\,K~\cite{whitfield_NeutronDif_2016}. 
According to our results, the delta phases are thermodynamically more stable than this orthorhombic phase at 0~K, 
and hence will become accessible at sufficiently low temperatures. 
However, the delta phases might be difficult to access since 
(a) the transition from orthorhombic symmetry to the delta phases is presumably 
a first-order transition (i.e., a discontinuous transition) 
based on connecting basins of symmetry~\cite{toledano1987landau,howard1998group} 
and (b) involves  significant structural rearrangement of 
internal coordinates to form the relevant delta motifs with a presumably kinetic activation barrier 
that is unlikely to be overcome below 10\,K. 
 
An issue that demands further investigation is whether the inclusion of zero-point energy (ZPE) and vibrational entropy 
influences the energetic ordering, i.e., if there is a strong entropic penalty of the delta phases with respect to the orthorhombic structure.
However, addressing this issue is challenging, since both harmonic and quasi-harmonic phonon calculations lead to imaginary modes that arise from ill described Van der Waals interactions and volume effects~\cite{Brivio_Frost_PRB_2015}. 
It has been also established that the rotational modes of the organic \ce{CH3NH3+} cations are strongly anharmonic, 
and that there is a relatively large anharmonic coupling between the organic librational modes and the
different optical modes of the octahedral framework~\cite{mamon-Giustino_anaharmonic_2017}. 
Indeed, vibrational entropy could in principle dominate the phase stability~\cite{thesame_not_thesame}, 
but obtaining a conclusive picture to fully understand this issue in pervoskites is not trivial. 

Further, the structures of the double-delta and delta phases 
are closely related to 
the CsPbI$_3$ and the Yellow-FAPbI$_3$ (formamidinium) phases, respectively~\cite{yi2016entropic,ma2017stable,eperon2015inorganic,lai2017structural,stoumpos2016structure}.
The effect of humidity on the degradation of perovskite (MAPI) solar cell performance also plays a significant role, and
has been thoroughly studied in the past~\cite{manser2016making}. 
Thanks to those studies, it is well accepted that in the presence of water, H$_2$O molecules 
deprotonates CH$_3$NH$_3^+$~\cite{frost2014atomistic}, thereby leading to product compounds with wider band gaps 
and substantially altered structural motifs, which are surprisingly similar to the double-delta phase 
(see Manser et al.~Ref.~\cite{manser2016making}).  

As discussed in detail in section~\ref{sec:pressure}, the enthalpy of the high temperature polymorphs (cubic and tetragonal) 
greatly increases with respect to the orthorhombic phase at high pressure (see Fig.~\ref{fig:PV}). 
This finding is in agreement with the experimental evidence that suggests a 
transformation to a disordered (amorphous) phase~\cite{jaffe2017pressure,capitani2016high}. 
The apparent reaction path suggests that, starting from the tetragonal or cubic phase, the pressure-induced 
amorphization occurs rapidly and is accelerated by entropic (temperature) effects. 
Szafra\'nski-Katrusiak reported that the tetragonal phase ($I4/mcm$) transforms via a first-order transition to a 
IV-phase with $Im3$ symmetry at 0.35\,GPa, and above 2.5\,GPa either to a phase V ($Im3$) 
or to an amorphous phase~\cite{szafranski2017photovoltaic}. 
Note that in order to correctly assign a crystal structure to the phases IV and V, a cell with 8 formula units 
(the same number of f.u. to describe the double-delta phase) is required. 
The same authors also list a range of possible sources of uncertainties that arise when performing single-crystal XRD 
diffraction measurements under pressure. In particular, the use of intense synchrotron beams for diffraction studies enhances 
potentially misleading effects: multiple scattering and higher-harmonics of $\lambda$ diffraction.  
Due to these technical difficulties, a plethora of different structures have been suggested that 
can be indexed to the same XRD spectrum, for instance $Imm2$, $Pnma$, $Im3$, $Immm$ or $Fmmm$
have been assigned to the phases IV and V, with a degree of disorder of iodide ions and distorted structural parameters~\cite{jaffe2016high}. 
Other experiments even suggest metalization and stabilization of phases under a considerable amount of pressure~\cite{jaffe2017pressure}. 

Based on our careful analysis of the available literature work, none of the experimental high-pressure experiments used the 
low-temperature orthorhombic phase as the starting material.
Instead, our simulated XRD patterns show that the tetragonal phases is present in the samples of Capitani~\cite{capitani2016high}, Jiang~\cite{Jiang_pressuremapbi_angew} and Jeffe~\cite{jaffe2017pressure} (See SI). Based on our calculations, we therefore suggest that either 
the delta or double delta phases could be synthesized by compressing precursor samples in the orthorhombic ($Pnma$) phase in low-temperature compression experiments. Since the orthorhombic phase is strongly destabilized upon compression, a transition towards the double-delta or delta phases will be rapidly favored with increasing pressure. 
Such an approach based on cold-compression is often overlooked in materials synthesis, 
but has been shown in the past to be a promising method for stabilizing elemental phases in carbon 
and phosphorous~\cite{mao2003bonding,maxi_2012-Zcarbon,Elemental_P-temperature_2017}.

Finally, we discuss the electronic properties of the \mapbi phases. 
If they could be synthesized in experiments, delta and double-delta both would exhibit very distinct electronic properties compared to any known phases. 
A corner-sharing octahedra (as in cubic, tetragonal and orthorhombic phases) promote a destabilization of the $s^2$ 
lone pairs on the M$^{2+}$-sites, leading to very dispersive valence and conduction bands 
(high mobility of charge carriers). Additionally, they lead to band structures that exhibit direct band-gaps 
within 1.5 -- 2.1\,eV~\cite{stoumpos2016structure}, a range that is suitable for many (opto-)electronic applications. 
Filip et al.~\cite{steric_gaps_NatCom2014} showed through an {\it in silico} study that 
the optical gap could be potentially tuned from the mid-infrared to the visible range via steric engineering. 
Even though this work is limited by the assumed Platonic model of \ce{PbI6}, a rather restricted configurational 
landscape for perovskites materials, it was shown to be more generally true~\cite{meloni2016valence} 
(delta and double delta models were not considered their works). 
For the (double) delta phases, the difference in the band gaps arises from the different connectivity or interoctahedral topology, namely edge- or face-sharing, where the lone pair is lowered in energy and leads to flatter electronic bands. 
Overall, accessing the new octahedral motifs observed in the (double) delta \ce{PbI6} would increase the design space to explore new organic and perovskite inorganic-frameworks. We suggest to further investigate synthetic routes based on compression and strain engineering to stabilize such structural features. 

\section{Conclusions} 

To summarize, we perform an {\it in-silico} exploration of the structural landscape of hybrid organic-inorganic  lead-iodide perovskites. 
We employ for the first time very large, realistic simulation cells with up to 96 atoms (8 formula units) for an exhaustive structural search, 
which allowed us to describe all the three known polymorphs on the same theoretical footing. 
From an exploration of more than 140,000 crystal structures, we identify novel low-energy phases with previously unobserved structural motifs for this system. The most promising candidates are the delta and double-delta phases, which are closely related to the reported 
yellow formamidinium non-perovskite phase and Cs lead iodides~\cite{ma2017stable}. 
According to our DFT calculations at 0\,K, these delta-phases in \mapbi are lower in energy than the experimentally observed orthorhombic phase. 
In contrast to any known phases of MAPI, the delta-phases exhibit edge- and surface sharing octahedral motifs, 
which in turn give rise to larger band-gaps than perovskite phases.
Different orientations of these octahedral moieties upon compression lead to changes in the band gaps. 
Hence, the electronic structure can be readily affected through phase selection and strain engineering, e.g., in thin-film synthesis.

\section{Methods}

Unbiased configurational exploration of \mapbi was performed with the minima hopping algorithm~\cite{MHM,Amsler_2010}. 
Energy and forces at the forcefield level were evaluated using the General Utility Lattice Program (GULP) molecular modeling package~\cite{curtin_gulpref}, employing forcefield parameters developed by Handley and Freeman~\cite{handley_gulppotential}. 
Exploratory runs were conducted with simulations cells containing 12, 24, 48, and 96 atoms at zero pressure. 

Energies, atomic forces and stresses at the density-functional theory level were calculated with the Vienna Ab Initio Simulation Package~{\sc vasp}~\cite{VASP_Kresse}. Various approximations to the exchange-correlation functionals were tested, but the thermodynamic properties were computed using the strongly constrained and appropriately normed semi-local density functional meta-GGA (SCAN)~\cite{sun2016accurate}. 
A plane wave basis-set with a cutoff energy of 800\,eV was used within the projector augmented wave (PAW) formalism.
The basis set was constructed with Pb d-shell (14), I (7), N (5), C (4), H (1) electrons as valence states in the PAW potentials. 
All calculations were spin unpolarized. The reciprocal space was sampled using $\Gamma-$centered $k$-grid meshes with spacings of 0.3 \AA$^{-1}$. 
Geometry relaxations were performed with tight tolerance parameters, 
leading to atomic forces of less than 5\,meV/\AA, and stresses below 0.1\,eV/\AA$^3$.  

\section{acknowledgement}

D.T and S.G thanks C. Handley for his help with the \mapbi forcefield. 
M.A. acknowledges support from the Novartis Universit{\"a}t Basel Excellence Scholarship for Life Sciences 
and the Swiss National Science Foundation (Project No.\ P300P2-158407, P300P2-174475).
Computational resources from the Swiss National Supercomputing Center (CSCS) in Lugano are gratefully acknowledged,  
J.A.F.-L. for project s752, 
D.T. and S.G. for project s707, 
M.A. for project s700 and 
U.R. and A.B. for projects s789 and s672. 
This research was supported by the NCCR MARVEL, funded by the Swiss National Science Foundation.

\bibliographystyle{apsrev4-1}
\bibliography{MAPBI}

\begin{thebibliography}{89}%
\makeatletter
\providecommand \@ifxundefined [1]{%
 \@ifx{#1\undefined}
}%
\providecommand \@ifnum [1]{%
 \ifnum #1\expandafter \@firstoftwo
 \else \expandafter \@secondoftwo
 \fi
}%
\providecommand \@ifx [1]{%
 \ifx #1\expandafter \@firstoftwo
 \else \expandafter \@secondoftwo
 \fi
}%
\providecommand \natexlab [1]{#1}%
\providecommand \enquote  [1]{``#1''}%
\providecommand \bibnamefont  [1]{#1}%
\providecommand \bibfnamefont [1]{#1}%
\providecommand \citenamefont [1]{#1}%
\providecommand \href@noop [0]{\@secondoftwo}%
\providecommand \href [0]{\begingroup \@sanitize@url \@href}%
\providecommand \@href[1]{\@@startlink{#1}\@@href}%
\providecommand \@@href[1]{\endgroup#1\@@endlink}%
\providecommand \@sanitize@url [0]{\catcode `\\12\catcode `\$12\catcode
  `\&12\catcode `\#12\catcode `\^12\catcode `\_12\catcode `\%12\relax}%
\providecommand \@@startlink[1]{}%
\providecommand \@@endlink[0]{}%
\providecommand \url  [0]{\begingroup\@sanitize@url \@url }%
\providecommand \@url [1]{\endgroup\@href {#1}{\urlprefix }}%
\providecommand \urlprefix  [0]{URL }%
\providecommand \Eprint [0]{\href }%
\providecommand \doibase [0]{http://dx.doi.org/}%
\providecommand \selectlanguage [0]{\@gobble}%
\providecommand \bibinfo  [0]{\@secondoftwo}%
\providecommand \bibfield  [0]{\@secondoftwo}%
\providecommand \translation [1]{[#1]}%
\providecommand \BibitemOpen [0]{}%
\providecommand \bibitemStop [0]{}%
\providecommand \bibitemNoStop [0]{.\EOS\space}%
\providecommand \EOS [0]{\spacefactor3000\relax}%
\providecommand \BibitemShut  [1]{\csname bibitem#1\endcsname}%
\let\auto@bib@innerbib\@empty
\bibitem [{\citenamefont {Buriak}\ \emph {et~al.}(2017)\citenamefont {Buriak},
  \citenamefont {Kamat}, \citenamefont {Schanze}, \citenamefont {Alivisatos},
  \citenamefont {Murphy}, \citenamefont {Schatz}, \citenamefont {Scholes},
  \citenamefont {Stang},\ and\ \citenamefont
  {Weiss}}]{alivisatos_god-perovska2017}%
  \BibitemOpen
  \bibfield  {author} {\bibinfo {author} {\bibfnamefont {J.~M.}\ \bibnamefont
  {Buriak}}, \bibinfo {author} {\bibfnamefont {P.~V.}\ \bibnamefont {Kamat}},
  \bibinfo {author} {\bibfnamefont {K.~S.}\ \bibnamefont {Schanze}}, \bibinfo
  {author} {\bibfnamefont {A.~P.}\ \bibnamefont {Alivisatos}}, \bibinfo
  {author} {\bibfnamefont {C.~J.}\ \bibnamefont {Murphy}}, \bibinfo {author}
  {\bibfnamefont {G.~C.}\ \bibnamefont {Schatz}}, \bibinfo {author}
  {\bibfnamefont {G.~D.}\ \bibnamefont {Scholes}}, \bibinfo {author}
  {\bibfnamefont {P.~J.}\ \bibnamefont {Stang}}, \ and\ \bibinfo {author}
  {\bibfnamefont {P.~S.}\ \bibnamefont {Weiss}},\ }\href@noop {} {\enquote
  {\bibinfo {title} {Virtual issue on metal-halide perovskite nanocrystals a
  bright future for optoelectronics},}\ } (\bibinfo {year} {2017})\BibitemShut
  {NoStop}%
\bibitem [{\citenamefont {Nishihata}\ \emph {et~al.}(2002)\citenamefont
  {Nishihata}, \citenamefont {Mizuki}, \citenamefont {Akao}, \citenamefont
  {Tanaka}, \citenamefont {Uenishi}, \citenamefont {Kimura}, \citenamefont
  {Okamoto},\ and\ \citenamefont {Hamada}}]{10.1038/nature00893}%
  \BibitemOpen
  \bibfield  {author} {\bibinfo {author} {\bibfnamefont {Y.}~\bibnamefont
  {Nishihata}}, \bibinfo {author} {\bibfnamefont {J.}~\bibnamefont {Mizuki}},
  \bibinfo {author} {\bibfnamefont {T.}~\bibnamefont {Akao}}, \bibinfo {author}
  {\bibfnamefont {H.}~\bibnamefont {Tanaka}}, \bibinfo {author} {\bibfnamefont
  {M.}~\bibnamefont {Uenishi}}, \bibinfo {author} {\bibfnamefont
  {M.}~\bibnamefont {Kimura}}, \bibinfo {author} {\bibfnamefont
  {T.}~\bibnamefont {Okamoto}}, \ and\ \bibinfo {author} {\bibfnamefont
  {N.}~\bibnamefont {Hamada}},\ }\href {\doibase 10.1038/nature00893}
  {\bibfield  {journal} {\bibinfo  {journal} {Nature}\ }\textbf {\bibinfo
  {volume} {418}},\ \bibinfo {pages} {164 } (\bibinfo {year}
  {2002})}\BibitemShut {NoStop}%
\bibitem [{\citenamefont {Deschler}\ \emph {et~al.}(2014)\citenamefont
  {Deschler}, \citenamefont {Price}, \citenamefont {Pathak}, \citenamefont
  {Klintberg}, \citenamefont {Jarausch}, \citenamefont {Higler}, \citenamefont
  {H\"uttner}, \citenamefont {Leijtens}, \citenamefont {Stranks}, \citenamefont
  {Snaith}, \citenamefont {Atat\"ure}, \citenamefont {Phillips},\ and\
  \citenamefont {Friend}}]{doi:10.1021/jz5005285}%
  \BibitemOpen
  \bibfield  {author} {\bibinfo {author} {\bibfnamefont {F.}~\bibnamefont
  {Deschler}}, \bibinfo {author} {\bibfnamefont {M.}~\bibnamefont {Price}},
  \bibinfo {author} {\bibfnamefont {S.}~\bibnamefont {Pathak}}, \bibinfo
  {author} {\bibfnamefont {L.~E.}\ \bibnamefont {Klintberg}}, \bibinfo {author}
  {\bibfnamefont {D.-D.}\ \bibnamefont {Jarausch}}, \bibinfo {author}
  {\bibfnamefont {R.}~\bibnamefont {Higler}}, \bibinfo {author} {\bibfnamefont
  {S.}~\bibnamefont {H\"uttner}}, \bibinfo {author} {\bibfnamefont
  {T.}~\bibnamefont {Leijtens}}, \bibinfo {author} {\bibfnamefont {S.~D.}\
  \bibnamefont {Stranks}}, \bibinfo {author} {\bibfnamefont {H.~J.}\
  \bibnamefont {Snaith}}, \bibinfo {author} {\bibfnamefont {M.}~\bibnamefont
  {Atat\"ure}}, \bibinfo {author} {\bibfnamefont {R.~T.}\ \bibnamefont
  {Phillips}}, \ and\ \bibinfo {author} {\bibfnamefont {R.~H.}\ \bibnamefont
  {Friend}},\ }\href {\doibase 10.1021/jz5005285} {\bibfield  {journal}
  {\bibinfo  {journal} {J. Phys. Chem. Lett.}\ }\textbf {\bibinfo {volume}
  {5}},\ \bibinfo {pages} {1421} (\bibinfo {year} {2014})}\BibitemShut
  {NoStop}%
\bibitem [{\citenamefont {Okuda}\ \emph {et~al.}(2001)\citenamefont {Okuda},
  \citenamefont {Nakanishi}, \citenamefont {Miyasaka},\ and\ \citenamefont
  {Tokura}}]{PhysRevB.63.113104}%
  \BibitemOpen
  \bibfield  {author} {\bibinfo {author} {\bibfnamefont {T.}~\bibnamefont
  {Okuda}}, \bibinfo {author} {\bibfnamefont {K.}~\bibnamefont {Nakanishi}},
  \bibinfo {author} {\bibfnamefont {S.}~\bibnamefont {Miyasaka}}, \ and\
  \bibinfo {author} {\bibfnamefont {Y.}~\bibnamefont {Tokura}},\ }\href
  {\doibase 10.1103/PhysRevB.63.113104} {\bibfield  {journal} {\bibinfo
  {journal} {Phys. Rev. B}\ }\textbf {\bibinfo {volume} {63}},\ \bibinfo
  {pages} {113104} (\bibinfo {year} {2001})}\BibitemShut {NoStop}%
\bibitem [{\citenamefont {Liu}\ \emph {et~al.}(2013)\citenamefont {Liu},
  \citenamefont {Johnston},\ and\ \citenamefont
  {Snaith}}]{10.1038/nature12509}%
  \BibitemOpen
  \bibfield  {author} {\bibinfo {author} {\bibfnamefont {M.}~\bibnamefont
  {Liu}}, \bibinfo {author} {\bibfnamefont {M.~B.}\ \bibnamefont {Johnston}}, \
  and\ \bibinfo {author} {\bibfnamefont {H.~J.}\ \bibnamefont {Snaith}},\
  }\href {\doibase 10.1038/nature12509} {\bibfield  {journal} {\bibinfo
  {journal} {Nature}\ }\textbf {\bibinfo {volume} {501}},\ \bibinfo {pages}
  {395 } (\bibinfo {year} {2013})}\BibitemShut {NoStop}%
\bibitem [{\citenamefont {Lee}\ \emph {et~al.}(2012)\citenamefont {Lee},
  \citenamefont {Teuscher}, \citenamefont {Miyasaka}, \citenamefont
  {Murakami},\ and\ \citenamefont {Snaith}}]{Lee02112012}%
  \BibitemOpen
  \bibfield  {author} {\bibinfo {author} {\bibfnamefont {M.~M.}\ \bibnamefont
  {Lee}}, \bibinfo {author} {\bibfnamefont {J.}~\bibnamefont {Teuscher}},
  \bibinfo {author} {\bibfnamefont {T.}~\bibnamefont {Miyasaka}}, \bibinfo
  {author} {\bibfnamefont {T.~N.}\ \bibnamefont {Murakami}}, \ and\ \bibinfo
  {author} {\bibfnamefont {H.~J.}\ \bibnamefont {Snaith}},\ }\href {\doibase
  10.1126/science.1228604} {\bibfield  {journal} {\bibinfo  {journal}
  {Science}\ }\textbf {\bibinfo {volume} {338}},\ \bibinfo {pages} {643 }
  (\bibinfo {year} {2012})}\BibitemShut {NoStop}%
\bibitem [{\citenamefont {Eerenstein}\ \emph {et~al.}(2006)\citenamefont
  {Eerenstein}, \citenamefont {Mathur},\ and\ \citenamefont
  {Scott}}]{10.1038/nature05023}%
  \BibitemOpen
  \bibfield  {author} {\bibinfo {author} {\bibfnamefont {W.}~\bibnamefont
  {Eerenstein}}, \bibinfo {author} {\bibfnamefont {N.~D.}\ \bibnamefont
  {Mathur}}, \ and\ \bibinfo {author} {\bibfnamefont {J.~F.}\ \bibnamefont
  {Scott}},\ }\href {\doibase 10.1038/nature05023} {\bibfield  {journal}
  {\bibinfo  {journal} {Nature}\ }\textbf {\bibinfo {volume} {442}},\ \bibinfo
  {pages} {759 } (\bibinfo {year} {2006})}\BibitemShut {NoStop}%
\bibitem [{\citenamefont {Hur}\ \emph {et~al.}(2004)\citenamefont {Hur},
  \citenamefont {Park}, \citenamefont {Sharma}, \citenamefont {Ahn},
  \citenamefont {Guha},\ and\ \citenamefont {Cheong}}]{10.1038/nature02572}%
  \BibitemOpen
  \bibfield  {author} {\bibinfo {author} {\bibfnamefont {N.}~\bibnamefont
  {Hur}}, \bibinfo {author} {\bibfnamefont {S.}~\bibnamefont {Park}}, \bibinfo
  {author} {\bibfnamefont {P.~A.}\ \bibnamefont {Sharma}}, \bibinfo {author}
  {\bibfnamefont {J.~S.}\ \bibnamefont {Ahn}}, \bibinfo {author} {\bibfnamefont
  {S.}~\bibnamefont {Guha}}, \ and\ \bibinfo {author} {\bibfnamefont {S.-W.}\
  \bibnamefont {Cheong}},\ }\href {\doibase 10.1038/nature02572} {\bibfield
  {journal} {\bibinfo  {journal} {Nature}\ }\textbf {\bibinfo {volume} {429}},\
  \bibinfo {pages} {392 } (\bibinfo {year} {2004})}\BibitemShut {NoStop}%
\bibitem [{\citenamefont {Ramirez}(1997)}]{cmr}%
  \BibitemOpen
  \bibfield  {author} {\bibinfo {author} {\bibfnamefont {A.~P.}\ \bibnamefont
  {Ramirez}},\ }\href {\doibase 10.1088/0953-8984/9/39/005} {\bibfield
  {journal} {\bibinfo  {journal} {J. Phys: Condens. Matter}\ }\textbf {\bibinfo
  {volume} {9}},\ \bibinfo {pages} {8171 } (\bibinfo {year}
  {1997})}\BibitemShut {NoStop}%
\bibitem [{\citenamefont {{Beno}}\ \emph {et~al.}(1987)\citenamefont {{Beno}},
  \citenamefont {{Soderholm}}, \citenamefont {{Capone}}, \citenamefont
  {{Hinks}},\ and\ \citenamefont {{Jorgensen}}}]{1987ApPhL..51...57B}%
  \BibitemOpen
  \bibfield  {author} {\bibinfo {author} {\bibfnamefont {M.~A.}\ \bibnamefont
  {{Beno}}}, \bibinfo {author} {\bibfnamefont {L.}~\bibnamefont {{Soderholm}}},
  \bibinfo {author} {\bibfnamefont {D.~W.}\ \bibnamefont {{Capone}},
  \bibfnamefont {II}}, \bibinfo {author} {\bibfnamefont {D.~G.}\ \bibnamefont
  {{Hinks}}}, \ and\ \bibinfo {author} {\bibfnamefont {J.~D.}\ \bibnamefont
  {{Jorgensen}}},\ }\href {\doibase 10.1063/1.98886} {\bibfield  {journal}
  {\bibinfo  {journal} {Appl. Phys. Lett.}\ }\textbf {\bibinfo {volume} {51}},\
  \bibinfo {pages} {57 } (\bibinfo {year} {1987})}\BibitemShut {NoStop}%
\bibitem [{\citenamefont {Wu}\ \emph {et~al.}(1987)\citenamefont {Wu},
  \citenamefont {Ashburn}, \citenamefont {Torng}, \citenamefont {Hor},
  \citenamefont {Meng}, \citenamefont {Gao}, \citenamefont {Huang},
  \citenamefont {Wang},\ and\ \citenamefont {Chu}}]{PhysRevLett.58.908}%
  \BibitemOpen
  \bibfield  {author} {\bibinfo {author} {\bibfnamefont {M.~K.}\ \bibnamefont
  {Wu}}, \bibinfo {author} {\bibfnamefont {J.~R.}\ \bibnamefont {Ashburn}},
  \bibinfo {author} {\bibfnamefont {C.~J.}\ \bibnamefont {Torng}}, \bibinfo
  {author} {\bibfnamefont {P.~H.}\ \bibnamefont {Hor}}, \bibinfo {author}
  {\bibfnamefont {R.~L.}\ \bibnamefont {Meng}}, \bibinfo {author}
  {\bibfnamefont {L.}~\bibnamefont {Gao}}, \bibinfo {author} {\bibfnamefont
  {Z.~J.}\ \bibnamefont {Huang}}, \bibinfo {author} {\bibfnamefont {Y.~Q.}\
  \bibnamefont {Wang}}, \ and\ \bibinfo {author} {\bibfnamefont {C.~W.}\
  \bibnamefont {Chu}},\ }\href {\doibase 10.1103/PhysRevLett.58.908} {\bibfield
   {journal} {\bibinfo  {journal} {Phys. Rev. Lett.}\ }\textbf {\bibinfo
  {volume} {58}},\ \bibinfo {pages} {908 } (\bibinfo {year}
  {1987})}\BibitemShut {NoStop}%
\bibitem [{\citenamefont {Haruyama}\ \emph {et~al.}(2014)\citenamefont
  {Haruyama}, \citenamefont {Sodeyama}, \citenamefont {Han},\ and\
  \citenamefont {Tateyama}}]{Tateyama_termination_2014}%
  \BibitemOpen
  \bibfield  {author} {\bibinfo {author} {\bibfnamefont {J.}~\bibnamefont
  {Haruyama}}, \bibinfo {author} {\bibfnamefont {K.}~\bibnamefont {Sodeyama}},
  \bibinfo {author} {\bibfnamefont {L.}~\bibnamefont {Han}}, \ and\ \bibinfo
  {author} {\bibfnamefont {Y.}~\bibnamefont {Tateyama}},\ }\href {\doibase
  10.1021/jz501510v} {\bibfield  {journal} {\bibinfo  {journal} {The Journal of
  Physical Chemistry Letters}\ }\textbf {\bibinfo {volume} {5}},\ \bibinfo
  {pages} {2903} (\bibinfo {year} {2014})},\ \bibinfo {note} {pMID: 26278097},\
  \Eprint {http://arxiv.org/abs/https://doi.org/10.1021/jz501510v}
  {https://doi.org/10.1021/jz501510v} \BibitemShut {NoStop}%
\bibitem [{rep()}]{report-piezoelectrics}%
  \BibitemOpen
  \href@noop {} {}\bibinfo {note} {Innovative Research and Products (iRAP) Inc.
  ET112: Piezoelectric Actuators and Motors-Types, Applications, new
  developments, Industry Structure and Global Markets (2010).}\BibitemShut
  {Stop}%
\bibitem [{\citenamefont {Mitchell}(2002)}]{mitchell2002perovskites}%
  \BibitemOpen
  \bibfield  {author} {\bibinfo {author} {\bibfnamefont {R.~H.}\ \bibnamefont
  {Mitchell}},\ }\href@noop {} {\emph {\bibinfo {title} {Perovskites: modern
  and ancient}}}\ (\bibinfo  {publisher} {Almaz Press Thunder Bay},\ \bibinfo
  {year} {2002})\BibitemShut {NoStop}%
\bibitem [{\citenamefont {Kojima}\ \emph {et~al.}(2009)\citenamefont {Kojima},
  \citenamefont {Teshima}, \citenamefont {Shirai},\ and\ \citenamefont
  {Miyasaka}}]{kojima_JACS}%
  \BibitemOpen
  \bibfield  {author} {\bibinfo {author} {\bibfnamefont {A.}~\bibnamefont
  {Kojima}}, \bibinfo {author} {\bibfnamefont {K.}~\bibnamefont {Teshima}},
  \bibinfo {author} {\bibfnamefont {Y.}~\bibnamefont {Shirai}}, \ and\ \bibinfo
  {author} {\bibfnamefont {T.}~\bibnamefont {Miyasaka}},\ }\href {\doibase
  10.1021/ja809598r} {\bibfield  {journal} {\bibinfo  {journal} {Journal of the
  American Chemical Society}\ }\textbf {\bibinfo {volume} {131}},\ \bibinfo
  {pages} {6050} (\bibinfo {year} {2009})},\ \Eprint
  {http://arxiv.org/abs/http://dx.doi.org/10.1021/ja809598r}
  {http://dx.doi.org/10.1021/ja809598r} \BibitemShut {NoStop}%
\bibitem [{\citenamefont {Oba}\ and\ \citenamefont
  {Kumagai}(2018)}]{oba2018design}%
  \BibitemOpen
  \bibfield  {author} {\bibinfo {author} {\bibfnamefont {F.}~\bibnamefont
  {Oba}}\ and\ \bibinfo {author} {\bibfnamefont {Y.}~\bibnamefont {Kumagai}},\
  }\href@noop {} {\bibfield  {journal} {\bibinfo  {journal} {Applied Physics
  Express}\ }\textbf {\bibinfo {volume} {11}},\ \bibinfo {pages} {060101}
  (\bibinfo {year} {2018})}\BibitemShut {NoStop}%
\bibitem [{\citenamefont {Lee}\ \emph {et~al.}(2005)\citenamefont {Lee},
  \citenamefont {Christen}, \citenamefont {Chisholm}, \citenamefont {Rouleau},\
  and\ \citenamefont {Lowndes}}]{lee2005strong}%
  \BibitemOpen
  \bibfield  {author} {\bibinfo {author} {\bibfnamefont {H.~N.}\ \bibnamefont
  {Lee}}, \bibinfo {author} {\bibfnamefont {H.~M.}\ \bibnamefont {Christen}},
  \bibinfo {author} {\bibfnamefont {M.~F.}\ \bibnamefont {Chisholm}}, \bibinfo
  {author} {\bibfnamefont {C.~M.}\ \bibnamefont {Rouleau}}, \ and\ \bibinfo
  {author} {\bibfnamefont {D.~H.}\ \bibnamefont {Lowndes}},\ }\href@noop {}
  {\bibfield  {journal} {\bibinfo  {journal} {Nature}\ }\textbf {\bibinfo
  {volume} {433}},\ \bibinfo {pages} {395} (\bibinfo {year}
  {2005})}\BibitemShut {NoStop}%
\bibitem [{\citenamefont {Whitfield}\ \emph {et~al.}(2016)\citenamefont
  {Whitfield}, \citenamefont {Herron}, \citenamefont {Guise}, \citenamefont
  {Page}, \citenamefont {Cheng}, \citenamefont {Milas},\ and\ \citenamefont
  {Crawford}}]{whitfield_NeutronDif_2016}%
  \BibitemOpen
  \bibfield  {author} {\bibinfo {author} {\bibfnamefont {P.}~\bibnamefont
  {Whitfield}}, \bibinfo {author} {\bibfnamefont {N.}~\bibnamefont {Herron}},
  \bibinfo {author} {\bibfnamefont {W.}~\bibnamefont {Guise}}, \bibinfo
  {author} {\bibfnamefont {K.}~\bibnamefont {Page}}, \bibinfo {author}
  {\bibfnamefont {Y.}~\bibnamefont {Cheng}}, \bibinfo {author} {\bibfnamefont
  {I.}~\bibnamefont {Milas}}, \ and\ \bibinfo {author} {\bibfnamefont
  {M.}~\bibnamefont {Crawford}},\ }\href@noop {} {\bibfield  {journal}
  {\bibinfo  {journal} {Scientific reports}\ }\textbf {\bibinfo {volume} {6}},\
  \bibinfo {pages} {35685} (\bibinfo {year} {2016})}\BibitemShut {NoStop}%
\bibitem [{\citenamefont {Glazer}(1972)}]{glazer_notation_1972}%
  \BibitemOpen
  \bibfield  {author} {\bibinfo {author} {\bibfnamefont {A.}~\bibnamefont
  {Glazer}},\ }\href@noop {} {\bibfield  {journal} {\bibinfo  {journal} {Acta
  Crystallographica Section B: Structural Crystallography and Crystal
  Chemistry}\ }\textbf {\bibinfo {volume} {28}},\ \bibinfo {pages} {3384}
  (\bibinfo {year} {1972})}\BibitemShut {NoStop}%
\bibitem [{\citenamefont {Capitani}\ \emph {et~al.}(2016)\citenamefont
  {Capitani}, \citenamefont {Marini}, \citenamefont {Caramazza}, \citenamefont
  {Postorino}, \citenamefont {Garbarino}, \citenamefont {Hanfland},
  \citenamefont {Pisanu}, \citenamefont {Quadrelli},\ and\ \citenamefont
  {Malavasi}}]{capitani2016high}%
  \BibitemOpen
  \bibfield  {author} {\bibinfo {author} {\bibfnamefont {F.}~\bibnamefont
  {Capitani}}, \bibinfo {author} {\bibfnamefont {C.}~\bibnamefont {Marini}},
  \bibinfo {author} {\bibfnamefont {S.}~\bibnamefont {Caramazza}}, \bibinfo
  {author} {\bibfnamefont {P.}~\bibnamefont {Postorino}}, \bibinfo {author}
  {\bibfnamefont {G.}~\bibnamefont {Garbarino}}, \bibinfo {author}
  {\bibfnamefont {M.}~\bibnamefont {Hanfland}}, \bibinfo {author}
  {\bibfnamefont {A.}~\bibnamefont {Pisanu}}, \bibinfo {author} {\bibfnamefont
  {P.}~\bibnamefont {Quadrelli}}, \ and\ \bibinfo {author} {\bibfnamefont
  {L.}~\bibnamefont {Malavasi}},\ }\href@noop {} {\bibfield  {journal}
  {\bibinfo  {journal} {Journal of Applied Physics}\ }\textbf {\bibinfo
  {volume} {119}},\ \bibinfo {pages} {185901} (\bibinfo {year}
  {2016})}\BibitemShut {NoStop}%
\bibitem [{\citenamefont {Szafranski}\ and\ \citenamefont
  {Katrusiak}(2016)}]{szafranski2016mechanism}%
  \BibitemOpen
  \bibfield  {author} {\bibinfo {author} {\bibfnamefont {M.}~\bibnamefont
  {Szafranski}}\ and\ \bibinfo {author} {\bibfnamefont {A.}~\bibnamefont
  {Katrusiak}},\ }\href@noop {} {\bibfield  {journal} {\bibinfo  {journal} {The
  journal of physical chemistry letters}\ }\textbf {\bibinfo {volume} {7}},\
  \bibinfo {pages} {3458} (\bibinfo {year} {2016})}\BibitemShut {NoStop}%
\bibitem [{\citenamefont {Szafra{\'n}ski}\ and\ \citenamefont
  {Katrusiak}(2017)}]{szafranski2017photovoltaic}%
  \BibitemOpen
  \bibfield  {author} {\bibinfo {author} {\bibfnamefont {M.}~\bibnamefont
  {Szafra{\'n}ski}}\ and\ \bibinfo {author} {\bibfnamefont {A.}~\bibnamefont
  {Katrusiak}},\ }\href@noop {} {\bibfield  {journal} {\bibinfo  {journal} {The
  journal of physical chemistry letters}\ }\textbf {\bibinfo {volume} {8}},\
  \bibinfo {pages} {2496} (\bibinfo {year} {2017})}\BibitemShut {NoStop}%
\bibitem [{\citenamefont {Jiang}\ \emph {et~al.}(2016)\citenamefont {Jiang},
  \citenamefont {Fang}, \citenamefont {Li}, \citenamefont {Xiao}, \citenamefont
  {Crowley}, \citenamefont {Wang}, \citenamefont {White}, \citenamefont
  {Goddard}, \citenamefont {Wang}, \citenamefont {Baikie},\ and\ \citenamefont
  {Fang}}]{Jiang_pressuremapbi_angew}%
  \BibitemOpen
  \bibfield  {author} {\bibinfo {author} {\bibfnamefont {S.}~\bibnamefont
  {Jiang}}, \bibinfo {author} {\bibfnamefont {Y.}~\bibnamefont {Fang}},
  \bibinfo {author} {\bibfnamefont {R.}~\bibnamefont {Li}}, \bibinfo {author}
  {\bibfnamefont {H.}~\bibnamefont {Xiao}}, \bibinfo {author} {\bibfnamefont
  {J.}~\bibnamefont {Crowley}}, \bibinfo {author} {\bibfnamefont
  {C.}~\bibnamefont {Wang}}, \bibinfo {author} {\bibfnamefont {T.~J.}\
  \bibnamefont {White}}, \bibinfo {author} {\bibfnamefont {W.~A.}\ \bibnamefont
  {Goddard}}, \bibinfo {author} {\bibfnamefont {Z.}~\bibnamefont {Wang}},
  \bibinfo {author} {\bibfnamefont {T.}~\bibnamefont {Baikie}}, \ and\ \bibinfo
  {author} {\bibfnamefont {J.}~\bibnamefont {Fang}},\ }\href {\doibase
  10.1002/anie.201601788} {\bibfield  {journal} {\bibinfo  {journal}
  {Angewandte Chemie International Edition}\ }\textbf {\bibinfo {volume}
  {55}},\ \bibinfo {pages} {6540} (\bibinfo {year} {2016})}\BibitemShut
  {NoStop}%
\bibitem [{\citenamefont {Jaffe}\ \emph {et~al.}(2016)\citenamefont {Jaffe},
  \citenamefont {Lin}, \citenamefont {Beavers}, \citenamefont {Voss},
  \citenamefont {Mao},\ and\ \citenamefont {Karunadasa}}]{jaffe2016high}%
  \BibitemOpen
  \bibfield  {author} {\bibinfo {author} {\bibfnamefont {A.}~\bibnamefont
  {Jaffe}}, \bibinfo {author} {\bibfnamefont {Y.}~\bibnamefont {Lin}}, \bibinfo
  {author} {\bibfnamefont {C.~M.}\ \bibnamefont {Beavers}}, \bibinfo {author}
  {\bibfnamefont {J.}~\bibnamefont {Voss}}, \bibinfo {author} {\bibfnamefont
  {W.~L.}\ \bibnamefont {Mao}}, \ and\ \bibinfo {author} {\bibfnamefont
  {H.~I.}\ \bibnamefont {Karunadasa}},\ }\href@noop {} {\bibfield  {journal}
  {\bibinfo  {journal} {ACS central science}\ }\textbf {\bibinfo {volume}
  {2}},\ \bibinfo {pages} {201} (\bibinfo {year} {2016})}\BibitemShut {NoStop}%
\bibitem [{\citenamefont {Jaffe}\ \emph
  {et~al.}(2017{\natexlab{a}})\citenamefont {Jaffe}, \citenamefont {Lin},\ and\
  \citenamefont {Karunadasa}}]{jaffe2017halide}%
  \BibitemOpen
  \bibfield  {author} {\bibinfo {author} {\bibfnamefont {A.}~\bibnamefont
  {Jaffe}}, \bibinfo {author} {\bibfnamefont {Y.}~\bibnamefont {Lin}}, \ and\
  \bibinfo {author} {\bibfnamefont {H.~I.}\ \bibnamefont {Karunadasa}},\
  }\href@noop {} {\bibfield  {journal} {\bibinfo  {journal} {ACS Energy
  Letters}\ }\textbf {\bibinfo {volume} {2}},\ \bibinfo {pages} {1549}
  (\bibinfo {year} {2017}{\natexlab{a}})}\BibitemShut {NoStop}%
\bibitem [{\citenamefont {Even}\ \emph {et~al.}(2013)\citenamefont {Even},
  \citenamefont {Pedesseau}, \citenamefont {Jancu},\ and\ \citenamefont
  {Katan}}]{even2013importance}%
  \BibitemOpen
  \bibfield  {author} {\bibinfo {author} {\bibfnamefont {J.}~\bibnamefont
  {Even}}, \bibinfo {author} {\bibfnamefont {L.}~\bibnamefont {Pedesseau}},
  \bibinfo {author} {\bibfnamefont {J.-M.}\ \bibnamefont {Jancu}}, \ and\
  \bibinfo {author} {\bibfnamefont {C.}~\bibnamefont {Katan}},\ }\href@noop {}
  {\bibfield  {journal} {\bibinfo  {journal} {The Journal of Physical Chemistry
  Letters}\ }\textbf {\bibinfo {volume} {4}},\ \bibinfo {pages} {2999}
  (\bibinfo {year} {2013})}\BibitemShut {NoStop}%
\bibitem [{\citenamefont {Korbel}\ \emph {et~al.}(2018)\citenamefont {Korbel},
  \citenamefont {Botti},\ and\ \citenamefont {Marques}}]{korbel2018stable}%
  \BibitemOpen
  \bibfield  {author} {\bibinfo {author} {\bibfnamefont {S.~M.}\ \bibnamefont
  {Korbel}}, \bibinfo {author} {\bibfnamefont {S.}~\bibnamefont {Botti}}, \
  and\ \bibinfo {author} {\bibfnamefont {M.~A.}\ \bibnamefont {Marques}},\
  }\href@noop {} {\bibfield  {journal} {\bibinfo  {journal} {Journal of
  Materials Chemistry A}\ } (\bibinfo {year} {2018})}\BibitemShut {NoStop}%
\bibitem [{\citenamefont {Emery}\ \emph {et~al.}(2016)\citenamefont {Emery},
  \citenamefont {Saal}, \citenamefont {Kirklin}, \citenamefont {Hegde},\ and\
  \citenamefont {Wolverton}}]{emery2016high}%
  \BibitemOpen
  \bibfield  {author} {\bibinfo {author} {\bibfnamefont {A.~A.}\ \bibnamefont
  {Emery}}, \bibinfo {author} {\bibfnamefont {J.~E.}\ \bibnamefont {Saal}},
  \bibinfo {author} {\bibfnamefont {S.}~\bibnamefont {Kirklin}}, \bibinfo
  {author} {\bibfnamefont {V.~I.}\ \bibnamefont {Hegde}}, \ and\ \bibinfo
  {author} {\bibfnamefont {C.}~\bibnamefont {Wolverton}},\ }\href@noop {}
  {\bibfield  {journal} {\bibinfo  {journal} {Chem. Mater.}\ }\textbf {\bibinfo
  {volume} {28}},\ \bibinfo {pages} {5621} (\bibinfo {year}
  {2016})}\BibitemShut {NoStop}%
\bibitem [{\citenamefont {Schmidt}\ \emph {et~al.}(2017)\citenamefont
  {Schmidt}, \citenamefont {Shi}, \citenamefont {Borlido}, \citenamefont
  {Chen}, \citenamefont {Botti},\ and\ \citenamefont
  {Marques}}]{schmidt2017predicting}%
  \BibitemOpen
  \bibfield  {author} {\bibinfo {author} {\bibfnamefont {J.}~\bibnamefont
  {Schmidt}}, \bibinfo {author} {\bibfnamefont {J.}~\bibnamefont {Shi}},
  \bibinfo {author} {\bibfnamefont {P.}~\bibnamefont {Borlido}}, \bibinfo
  {author} {\bibfnamefont {L.}~\bibnamefont {Chen}}, \bibinfo {author}
  {\bibfnamefont {S.}~\bibnamefont {Botti}}, \ and\ \bibinfo {author}
  {\bibfnamefont {M.~A.}\ \bibnamefont {Marques}},\ }\href@noop {} {\bibfield
  {journal} {\bibinfo  {journal} {Chemistry of Materials}\ }\textbf {\bibinfo
  {volume} {29}},\ \bibinfo {pages} {5090} (\bibinfo {year}
  {2017})}\BibitemShut {NoStop}%
\bibitem [{\citenamefont {Sarmiento-Perez}\ \emph {et~al.}(2015)\citenamefont
  {Sarmiento-Perez}, \citenamefont {Cerqueira}, \citenamefont {K{\"o}rbel},
  \citenamefont {Botti},\ and\ \citenamefont
  {Marques}}]{sarmiento2015prediction}%
  \BibitemOpen
  \bibfield  {author} {\bibinfo {author} {\bibfnamefont {R.}~\bibnamefont
  {Sarmiento-Perez}}, \bibinfo {author} {\bibfnamefont {T.~F.}\ \bibnamefont
  {Cerqueira}}, \bibinfo {author} {\bibfnamefont {S.}~\bibnamefont
  {K{\"o}rbel}}, \bibinfo {author} {\bibfnamefont {S.}~\bibnamefont {Botti}}, \
  and\ \bibinfo {author} {\bibfnamefont {M.~A.}\ \bibnamefont {Marques}},\
  }\href@noop {} {\bibfield  {journal} {\bibinfo  {journal} {Chemistry of
  Materials}\ }\textbf {\bibinfo {volume} {27}},\ \bibinfo {pages} {5957}
  (\bibinfo {year} {2015})}\BibitemShut {NoStop}%
\bibitem [{\citenamefont {Huan}\ \emph {et~al.}(2016)\citenamefont {Huan},
  \citenamefont {Tuoc},\ and\ \citenamefont {Minh}}]{Huan_PhysRevB}%
  \BibitemOpen
  \bibfield  {author} {\bibinfo {author} {\bibfnamefont {T.~D.}\ \bibnamefont
  {Huan}}, \bibinfo {author} {\bibfnamefont {V.~N.}\ \bibnamefont {Tuoc}}, \
  and\ \bibinfo {author} {\bibfnamefont {N.~V.}\ \bibnamefont {Minh}},\ }\href
  {\doibase 10.1103/PhysRevB.93.094105} {\bibfield  {journal} {\bibinfo
  {journal} {Phys. Rev. B}\ }\textbf {\bibinfo {volume} {93}},\ \bibinfo
  {pages} {094105} (\bibinfo {year} {2016})}\BibitemShut {NoStop}%
\bibitem [{\citenamefont {Kim}\ \emph {et~al.}(2017{\natexlab{a}})\citenamefont
  {Kim}, \citenamefont {Huan}, \citenamefont {Krishnan},\ and\ \citenamefont
  {Ramprasad}}]{Kim_SciData}%
  \BibitemOpen
  \bibfield  {author} {\bibinfo {author} {\bibfnamefont {C.}~\bibnamefont
  {Kim}}, \bibinfo {author} {\bibfnamefont {T.~D.}\ \bibnamefont {Huan}},
  \bibinfo {author} {\bibfnamefont {S.}~\bibnamefont {Krishnan}}, \ and\
  \bibinfo {author} {\bibfnamefont {R.}~\bibnamefont {Ramprasad}},\ }\href
  {http://dx.doi.org/10.1038/sdata.2017.57} {\bibfield  {journal} {\bibinfo
  {journal} {Scientific Data}\ }\textbf {\bibinfo {volume} {4}},\ \bibinfo
  {pages} {170057 EP } (\bibinfo {year} {2017}{\natexlab{a}})},\ \bibinfo
  {note} {data Descriptor}\BibitemShut {NoStop}%
\bibitem [{\citenamefont {Goedecker}(2004)}]{MHM}%
  \BibitemOpen
  \bibfield  {author} {\bibinfo {author} {\bibfnamefont {S.}~\bibnamefont
  {Goedecker}},\ }\href@noop {} {\bibfield  {journal} {\bibinfo  {journal} {The
  Journal of Chemical Physics}\ }\textbf {\bibinfo {volume} {120}},\ \bibinfo
  {pages} {9911} (\bibinfo {year} {2004})}\BibitemShut {NoStop}%
\bibitem [{\citenamefont {Amsler}\ and\ \citenamefont
  {Goedecker}(2010)}]{Amsler_2010}%
  \BibitemOpen
  \bibfield  {author} {\bibinfo {author} {\bibfnamefont {M.}~\bibnamefont
  {Amsler}}\ and\ \bibinfo {author} {\bibfnamefont {S.}~\bibnamefont
  {Goedecker}},\ }\href@noop {} {\bibfield  {journal} {\bibinfo  {journal} {The
  Journal of Chemical Physics}\ }\textbf {\bibinfo {volume} {133}},\ \bibinfo
  {pages} {224104} (\bibinfo {year} {2010})}\BibitemShut {NoStop}%
\bibitem [{\citenamefont {Handley}\ and\ \citenamefont
  {Freeman}(2017)}]{handley_gulppotential}%
  \BibitemOpen
  \bibfield  {author} {\bibinfo {author} {\bibfnamefont {C.~M.}\ \bibnamefont
  {Handley}}\ and\ \bibinfo {author} {\bibfnamefont {C.~L.}\ \bibnamefont
  {Freeman}},\ }\href {\doibase 10.1039/C6CP05829A} {\bibfield  {journal}
  {\bibinfo  {journal} {Phys. Chem. Chem. Phys.}\ }\textbf {\bibinfo {volume}
  {19}},\ \bibinfo {pages} {2313} (\bibinfo {year} {2017})}\BibitemShut
  {NoStop}%
\bibitem [{\citenamefont {Gale}\ and\ \citenamefont
  {Rohl}(2003)}]{curtin_gulpref}%
  \BibitemOpen
  \bibfield  {author} {\bibinfo {author} {\bibfnamefont {J.~D.}\ \bibnamefont
  {Gale}}\ and\ \bibinfo {author} {\bibfnamefont {A.~L.}\ \bibnamefont
  {Rohl}},\ }\href {\doibase 10.1080/0892702031000104887} {\bibfield  {journal}
  {\bibinfo  {journal} {Molecular Simulation}\ }\textbf {\bibinfo {volume}
  {29}},\ \bibinfo {pages} {291} (\bibinfo {year} {2003})}\BibitemShut
  {NoStop}%
\bibitem [{\citenamefont {Perdew}\ and\ \citenamefont
  {Wang}(1992)}]{Perdew-Wang_1992}%
  \BibitemOpen
  \bibfield  {author} {\bibinfo {author} {\bibfnamefont {J.~P.}\ \bibnamefont
  {Perdew}}\ and\ \bibinfo {author} {\bibfnamefont {Y.}~\bibnamefont {Wang}},\
  }\href {\doibase 10.1103/PhysRevB.45.13244} {\bibfield  {journal} {\bibinfo
  {journal} {Phys. Rev. B}\ }\textbf {\bibinfo {volume} {45}},\ \bibinfo
  {pages} {13244} (\bibinfo {year} {1992})}\BibitemShut {NoStop}%
\bibitem [{\citenamefont {Perdew}\ \emph {et~al.}(1996)\citenamefont {Perdew},
  \citenamefont {Burke},\ and\ \citenamefont {Ernzerhof}}]{PBE96}%
  \BibitemOpen
  \bibfield  {author} {\bibinfo {author} {\bibfnamefont {J.~P.}\ \bibnamefont
  {Perdew}}, \bibinfo {author} {\bibfnamefont {K.}~\bibnamefont {Burke}}, \
  and\ \bibinfo {author} {\bibfnamefont {M.}~\bibnamefont {Ernzerhof}},\
  }\href@noop {} {\bibfield  {journal} {\bibinfo  {journal} {Phys. Rev. Lett.}\
  }\textbf {\bibinfo {volume} {77}},\ \bibinfo {pages} {3865} (\bibinfo {year}
  {1996})}\BibitemShut {NoStop}%
\bibitem [{\citenamefont {Perdew}\ \emph {et~al.}(2008)\citenamefont {Perdew},
  \citenamefont {Ruzsinszky}, \citenamefont {Csonka}, \citenamefont {Vydrov},
  \citenamefont {Scuseria}, \citenamefont {Constantin}, \citenamefont {Zhou},\
  and\ \citenamefont {Burke}}]{PBESol}%
  \BibitemOpen
  \bibfield  {author} {\bibinfo {author} {\bibfnamefont {J.~P.}\ \bibnamefont
  {Perdew}}, \bibinfo {author} {\bibfnamefont {A.}~\bibnamefont {Ruzsinszky}},
  \bibinfo {author} {\bibfnamefont {G.~I.}\ \bibnamefont {Csonka}}, \bibinfo
  {author} {\bibfnamefont {O.~A.}\ \bibnamefont {Vydrov}}, \bibinfo {author}
  {\bibfnamefont {G.~E.}\ \bibnamefont {Scuseria}}, \bibinfo {author}
  {\bibfnamefont {L.~A.}\ \bibnamefont {Constantin}}, \bibinfo {author}
  {\bibfnamefont {X.}~\bibnamefont {Zhou}}, \ and\ \bibinfo {author}
  {\bibfnamefont {K.}~\bibnamefont {Burke}},\ }\href@noop {} {\bibfield
  {journal} {\bibinfo  {journal} {Physical Review Letters}\ }\textbf {\bibinfo
  {volume} {100}},\ \bibinfo {pages} {136406} (\bibinfo {year}
  {2008})}\BibitemShut {NoStop}%
\bibitem [{\citenamefont {Grimme}(2006)}]{grimme2006_D2}%
  \BibitemOpen
  \bibfield  {author} {\bibinfo {author} {\bibfnamefont {S.}~\bibnamefont
  {Grimme}},\ }\href@noop {} {\bibfield  {journal} {\bibinfo  {journal}
  {Journal of computational chemistry}\ }\textbf {\bibinfo {volume} {27}},\
  \bibinfo {pages} {1787} (\bibinfo {year} {2006})}\BibitemShut {NoStop}%
\bibitem [{\citenamefont {Grimme}\ \emph {et~al.}(2010)\citenamefont {Grimme},
  \citenamefont {Antony}, \citenamefont {Ehrlich},\ and\ \citenamefont
  {Krieg}}]{grimme2010_D3}%
  \BibitemOpen
  \bibfield  {author} {\bibinfo {author} {\bibfnamefont {S.}~\bibnamefont
  {Grimme}}, \bibinfo {author} {\bibfnamefont {J.}~\bibnamefont {Antony}},
  \bibinfo {author} {\bibfnamefont {S.}~\bibnamefont {Ehrlich}}, \ and\
  \bibinfo {author} {\bibfnamefont {H.}~\bibnamefont {Krieg}},\ }\href@noop {}
  {\bibfield  {journal} {\bibinfo  {journal} {The Journal of chemical physics}\
  }\textbf {\bibinfo {volume} {132}},\ \bibinfo {pages} {154104} (\bibinfo
  {year} {2010})}\BibitemShut {NoStop}%
\bibitem [{\citenamefont {Tkatchenko}\ and\ \citenamefont
  {Scheffler}(2009)}]{tkatchenko_first_2009}%
  \BibitemOpen
  \bibfield  {author} {\bibinfo {author} {\bibfnamefont {A.}~\bibnamefont
  {Tkatchenko}}\ and\ \bibinfo {author} {\bibfnamefont {M.}~\bibnamefont
  {Scheffler}},\ }\href@noop {} {\bibfield  {journal} {\bibinfo  {journal}
  {Physical review letters}\ }\textbf {\bibinfo {volume} {102}},\ \bibinfo
  {pages} {073005} (\bibinfo {year} {2009})}\BibitemShut {NoStop}%
\bibitem [{\citenamefont {Tkatchenko}\ \emph {et~al.}(2012)\citenamefont
  {Tkatchenko}, \citenamefont {DiStasio~Jr}, \citenamefont {Car},\ and\
  \citenamefont {Scheffler}}]{manyBody_cor_tkatchenko2012}%
  \BibitemOpen
  \bibfield  {author} {\bibinfo {author} {\bibfnamefont {A.}~\bibnamefont
  {Tkatchenko}}, \bibinfo {author} {\bibfnamefont {R.~A.}\ \bibnamefont
  {DiStasio~Jr}}, \bibinfo {author} {\bibfnamefont {R.}~\bibnamefont {Car}}, \
  and\ \bibinfo {author} {\bibfnamefont {M.}~\bibnamefont {Scheffler}},\
  }\href@noop {} {\bibfield  {journal} {\bibinfo  {journal} {Physical review
  letters}\ }\textbf {\bibinfo {volume} {108}},\ \bibinfo {pages} {236402}
  (\bibinfo {year} {2012})}\BibitemShut {NoStop}%
\bibitem [{\citenamefont {Steinmann}\ and\ \citenamefont
  {Corminboeuf}(2011)}]{dDsC_vdW_2011}%
  \BibitemOpen
  \bibfield  {author} {\bibinfo {author} {\bibfnamefont {S.~N.}\ \bibnamefont
  {Steinmann}}\ and\ \bibinfo {author} {\bibfnamefont {C.}~\bibnamefont
  {Corminboeuf}},\ }\href@noop {} {\bibfield  {journal} {\bibinfo  {journal}
  {The Journal of chemical physics}\ }\textbf {\bibinfo {volume} {134}},\
  \bibinfo {pages} {044117} (\bibinfo {year} {2011})}\BibitemShut {NoStop}%
\bibitem [{\citenamefont {Adamo}\ and\ \citenamefont {Barone}(1999)}]{PBE0}%
  \BibitemOpen
  \bibfield  {author} {\bibinfo {author} {\bibfnamefont {C.}~\bibnamefont
  {Adamo}}\ and\ \bibinfo {author} {\bibfnamefont {V.}~\bibnamefont {Barone}},\
  }\href@noop {} {\bibfield  {journal} {\bibinfo  {journal} {J. Chem. Phys.}\
  }\textbf {\bibinfo {volume} {110}},\ \bibinfo {pages} {6158} (\bibinfo {year}
  {1999})}\BibitemShut {NoStop}%
\bibitem [{\citenamefont {Sun}\ \emph {et~al.}(2016{\natexlab{a}})\citenamefont
  {Sun}, \citenamefont {Remsing}, \citenamefont {Zhang}, \citenamefont {Sun},
  \citenamefont {Ruzsinszky}, \citenamefont {Peng}, \citenamefont {Yang},
  \citenamefont {Paul}, \citenamefont {Waghmare}, \citenamefont {Wu},
  \citenamefont {Klein},\ and\ \citenamefont {Perdew}}]{sun2016accurate}%
  \BibitemOpen
  \bibfield  {author} {\bibinfo {author} {\bibfnamefont {J.}~\bibnamefont
  {Sun}}, \bibinfo {author} {\bibfnamefont {R.~C.}\ \bibnamefont {Remsing}},
  \bibinfo {author} {\bibfnamefont {Y.}~\bibnamefont {Zhang}}, \bibinfo
  {author} {\bibfnamefont {Z.}~\bibnamefont {Sun}}, \bibinfo {author}
  {\bibfnamefont {A.}~\bibnamefont {Ruzsinszky}}, \bibinfo {author}
  {\bibfnamefont {H.}~\bibnamefont {Peng}}, \bibinfo {author} {\bibfnamefont
  {Z.}~\bibnamefont {Yang}}, \bibinfo {author} {\bibfnamefont {A.}~\bibnamefont
  {Paul}}, \bibinfo {author} {\bibfnamefont {U.}~\bibnamefont {Waghmare}},
  \bibinfo {author} {\bibfnamefont {X.}~\bibnamefont {Wu}}, \bibinfo {author}
  {\bibfnamefont {M.~L.}\ \bibnamefont {Klein}}, \ and\ \bibinfo {author}
  {\bibfnamefont {J.~P.}\ \bibnamefont {Perdew}},\ }\href@noop {} {\bibfield
  {journal} {\bibinfo  {journal} {Nature chemistry}\ }\textbf {\bibinfo
  {volume} {8}},\ \bibinfo {pages} {831} (\bibinfo {year}
  {2016}{\natexlab{a}})}\BibitemShut {NoStop}%
\bibitem [{\citenamefont {Li}\ and\ \citenamefont
  {Rinke}(2016)}]{Rinke_eggsVDW_2016}%
  \BibitemOpen
  \bibfield  {author} {\bibinfo {author} {\bibfnamefont {J.}~\bibnamefont
  {Li}}\ and\ \bibinfo {author} {\bibfnamefont {P.}~\bibnamefont {Rinke}},\
  }\href@noop {} {\bibfield  {journal} {\bibinfo  {journal} {Physical Review
  B}\ }\textbf {\bibinfo {volume} {94}},\ \bibinfo {pages} {045201} (\bibinfo
  {year} {2016})}\BibitemShut {NoStop}%
\bibitem [{\citenamefont {Wang}\ \emph {et~al.}(2014)\citenamefont {Wang},
  \citenamefont {Gould}, \citenamefont {Dobson}, \citenamefont {Zhang},
  \citenamefont {Yang}, \citenamefont {Yao},\ and\ \citenamefont
  {Zhao}}]{benchmark_pnma_2014}%
  \BibitemOpen
  \bibfield  {author} {\bibinfo {author} {\bibfnamefont {Y.}~\bibnamefont
  {Wang}}, \bibinfo {author} {\bibfnamefont {T.}~\bibnamefont {Gould}},
  \bibinfo {author} {\bibfnamefont {J.~F.}\ \bibnamefont {Dobson}}, \bibinfo
  {author} {\bibfnamefont {H.}~\bibnamefont {Zhang}}, \bibinfo {author}
  {\bibfnamefont {H.}~\bibnamefont {Yang}}, \bibinfo {author} {\bibfnamefont
  {X.}~\bibnamefont {Yao}}, \ and\ \bibinfo {author} {\bibfnamefont
  {H.}~\bibnamefont {Zhao}},\ }\href {\doibase 10.1039/C3CP54479F} {\bibfield
  {journal} {\bibinfo  {journal} {Phys. Chem. Chem. Phys.}\ }\textbf {\bibinfo
  {volume} {16}},\ \bibinfo {pages} {1424} (\bibinfo {year}
  {2014})}\BibitemShut {NoStop}%
\bibitem [{\citenamefont {P{\'e}rez-Osorio}\ \emph {et~al.}(2017)\citenamefont
  {P{\'e}rez-Osorio}, \citenamefont {Champagne}, \citenamefont {Zacharias},
  \citenamefont {Rignanese},\ and\ \citenamefont
  {Giustino}}]{mamon-Giustino_anaharmonic_2017}%
  \BibitemOpen
  \bibfield  {author} {\bibinfo {author} {\bibfnamefont {M.~A.}\ \bibnamefont
  {P{\'e}rez-Osorio}}, \bibinfo {author} {\bibfnamefont {A.}~\bibnamefont
  {Champagne}}, \bibinfo {author} {\bibfnamefont {M.}~\bibnamefont
  {Zacharias}}, \bibinfo {author} {\bibfnamefont {G.-M.}\ \bibnamefont
  {Rignanese}}, \ and\ \bibinfo {author} {\bibfnamefont {F.}~\bibnamefont
  {Giustino}},\ }\href@noop {} {\bibfield  {journal} {\bibinfo  {journal} {The
  Journal of Physical Chemistry C}\ }\textbf {\bibinfo {volume} {121}},\
  \bibinfo {pages} {18459} (\bibinfo {year} {2017})}\BibitemShut {NoStop}%
\bibitem [{\citenamefont {Wiktor}\ \emph {et~al.}(2017)\citenamefont {Wiktor},
  \citenamefont {Rothlisberger},\ and\ \citenamefont
  {Pasquarello}}]{wiktor2017predictive}%
  \BibitemOpen
  \bibfield  {author} {\bibinfo {author} {\bibfnamefont {J.}~\bibnamefont
  {Wiktor}}, \bibinfo {author} {\bibfnamefont {U.}~\bibnamefont
  {Rothlisberger}}, \ and\ \bibinfo {author} {\bibfnamefont {A.}~\bibnamefont
  {Pasquarello}},\ }\href@noop {} {\bibfield  {journal} {\bibinfo  {journal}
  {The Journal of Physical Chemistry Letters}\ }\textbf {\bibinfo {volume}
  {8}},\ \bibinfo {pages} {5507} (\bibinfo {year} {2017})}\BibitemShut
  {NoStop}%
\bibitem [{\citenamefont {Zhang}\ \emph {et~al.}(2018)\citenamefont {Zhang},
  \citenamefont {Kitchaev}, \citenamefont {Yang}, \citenamefont {Chen},
  \citenamefont {Dacek}, \citenamefont {Sarmiento-Perez}, \citenamefont
  {Marques}, \citenamefont {Peng}, \citenamefont {Ceder},\ and\ \citenamefont
  {Perdew}}]{scan_benchmark_2018}%
  \BibitemOpen
  \bibfield  {author} {\bibinfo {author} {\bibfnamefont {Y.}~\bibnamefont
  {Zhang}}, \bibinfo {author} {\bibfnamefont {D.~A.}\ \bibnamefont {Kitchaev}},
  \bibinfo {author} {\bibfnamefont {J.}~\bibnamefont {Yang}}, \bibinfo {author}
  {\bibfnamefont {T.}~\bibnamefont {Chen}}, \bibinfo {author} {\bibfnamefont
  {S.~T.}\ \bibnamefont {Dacek}}, \bibinfo {author} {\bibfnamefont {R.~A.}\
  \bibnamefont {Sarmiento-Perez}}, \bibinfo {author} {\bibfnamefont {M.~A.}\
  \bibnamefont {Marques}}, \bibinfo {author} {\bibfnamefont {H.}~\bibnamefont
  {Peng}}, \bibinfo {author} {\bibfnamefont {G.}~\bibnamefont {Ceder}}, \ and\
  \bibinfo {author} {\bibfnamefont {J.~P.}\ \bibnamefont {Perdew}},\
  }\href@noop {} {\bibfield  {journal} {\bibinfo  {journal} {npj Computational
  Materials}\ }\textbf {\bibinfo {volume} {4}},\ \bibinfo {pages} {9} (\bibinfo
  {year} {2018})}\BibitemShut {NoStop}%
\bibitem [{\citenamefont {Lahnsteiner}\ \emph {et~al.}(2016)\citenamefont
  {Lahnsteiner}, \citenamefont {Kresse}, \citenamefont {Kumar}, \citenamefont
  {Sarma}, \citenamefont {Franchini},\ and\ \citenamefont
  {Bokdam}}]{MD_Jonathan_PRB-2016}%
  \BibitemOpen
  \bibfield  {author} {\bibinfo {author} {\bibfnamefont {J.}~\bibnamefont
  {Lahnsteiner}}, \bibinfo {author} {\bibfnamefont {G.}~\bibnamefont {Kresse}},
  \bibinfo {author} {\bibfnamefont {A.}~\bibnamefont {Kumar}}, \bibinfo
  {author} {\bibfnamefont {D.~D.}\ \bibnamefont {Sarma}}, \bibinfo {author}
  {\bibfnamefont {C.}~\bibnamefont {Franchini}}, \ and\ \bibinfo {author}
  {\bibfnamefont {M.}~\bibnamefont {Bokdam}},\ }\href {\doibase
  10.1103/PhysRevB.94.214114} {\bibfield  {journal} {\bibinfo  {journal} {Phys.
  Rev. B}\ }\textbf {\bibinfo {volume} {94}},\ \bibinfo {pages} {214114}
  (\bibinfo {year} {2016})}\BibitemShut {NoStop}%
\bibitem [{\citenamefont {Bokdam}\ \emph {et~al.}(2017)\citenamefont {Bokdam},
  \citenamefont {Lahnsteiner}, \citenamefont {Ramberger}, \citenamefont
  {Sch\"afer},\ and\ \citenamefont
  {Kresse}}]{PRL_Menno-Jonathan_RPA-SCAN_2017}%
  \BibitemOpen
  \bibfield  {author} {\bibinfo {author} {\bibfnamefont {M.}~\bibnamefont
  {Bokdam}}, \bibinfo {author} {\bibfnamefont {J.}~\bibnamefont {Lahnsteiner}},
  \bibinfo {author} {\bibfnamefont {B.}~\bibnamefont {Ramberger}}, \bibinfo
  {author} {\bibfnamefont {T.}~\bibnamefont {Sch\"afer}}, \ and\ \bibinfo
  {author} {\bibfnamefont {G.}~\bibnamefont {Kresse}},\ }\href {\doibase
  10.1103/PhysRevLett.119.145501} {\bibfield  {journal} {\bibinfo  {journal}
  {Phys. Rev. Lett.}\ }\textbf {\bibinfo {volume} {119}},\ \bibinfo {pages}
  {145501} (\bibinfo {year} {2017})}\BibitemShut {NoStop}%
\bibitem [{\citenamefont {Zhu}\ \emph {et~al.}(2016)\citenamefont {Zhu},
  \citenamefont {Amsler}, \citenamefont {Fuhrer}, \citenamefont {Schaefer},
  \citenamefont {Faraji}, \citenamefont {Rostami}, \citenamefont {Ghasemi},
  \citenamefont {Sadeghi}, \citenamefont {Grauzinyte}, \citenamefont
  {Wolverton} \emph {et~al.}}]{zhu2016fingerprint}%
  \BibitemOpen
  \bibfield  {author} {\bibinfo {author} {\bibfnamefont {L.}~\bibnamefont
  {Zhu}}, \bibinfo {author} {\bibfnamefont {M.}~\bibnamefont {Amsler}},
  \bibinfo {author} {\bibfnamefont {T.}~\bibnamefont {Fuhrer}}, \bibinfo
  {author} {\bibfnamefont {B.}~\bibnamefont {Schaefer}}, \bibinfo {author}
  {\bibfnamefont {S.}~\bibnamefont {Faraji}}, \bibinfo {author} {\bibfnamefont
  {S.}~\bibnamefont {Rostami}}, \bibinfo {author} {\bibfnamefont {S.~A.}\
  \bibnamefont {Ghasemi}}, \bibinfo {author} {\bibfnamefont {A.}~\bibnamefont
  {Sadeghi}}, \bibinfo {author} {\bibfnamefont {M.}~\bibnamefont {Grauzinyte}},
  \bibinfo {author} {\bibfnamefont {C.}~\bibnamefont {Wolverton}},  \emph
  {et~al.},\ }\href@noop {} {\bibfield  {journal} {\bibinfo  {journal} {The
  Journal of chemical physics}\ }\textbf {\bibinfo {volume} {144}},\ \bibinfo
  {pages} {034203} (\bibinfo {year} {2016})}\BibitemShut {NoStop}%
\bibitem [{\citenamefont {Flores-Livas}\ \emph
  {et~al.}(2017{\natexlab{a}})\citenamefont {Flores-Livas}, \citenamefont
  {Sanna},\ and\ \citenamefont {Goedecker}}]{flores2017accelerated}%
  \BibitemOpen
  \bibfield  {author} {\bibinfo {author} {\bibfnamefont {J.~A.}\ \bibnamefont
  {Flores-Livas}}, \bibinfo {author} {\bibfnamefont {A.}~\bibnamefont {Sanna}},
  \ and\ \bibinfo {author} {\bibfnamefont {S.}~\bibnamefont {Goedecker}},\
  }\href@noop {} {\bibfield  {journal} {\bibinfo  {journal} {Novel
  Superconducting Materials}\ }\textbf {\bibinfo {volume} {3}},\ \bibinfo
  {pages} {6} (\bibinfo {year} {2017}{\natexlab{a}})}\BibitemShut {NoStop}%
\bibitem [{\citenamefont {Sun}\ \emph {et~al.}(2016{\natexlab{b}})\citenamefont
  {Sun}, \citenamefont {Dacek}, \citenamefont {Ong}, \citenamefont {Hautier},
  \citenamefont {Jain}, \citenamefont {Richards}, \citenamefont {Gamst},
  \citenamefont {Persson},\ and\ \citenamefont {Ceder}}]{Sun2016thermodynamic}%
  \BibitemOpen
  \bibfield  {author} {\bibinfo {author} {\bibfnamefont {W.}~\bibnamefont
  {Sun}}, \bibinfo {author} {\bibfnamefont {S.~T.}\ \bibnamefont {Dacek}},
  \bibinfo {author} {\bibfnamefont {S.~P.}\ \bibnamefont {Ong}}, \bibinfo
  {author} {\bibfnamefont {G.}~\bibnamefont {Hautier}}, \bibinfo {author}
  {\bibfnamefont {A.}~\bibnamefont {Jain}}, \bibinfo {author} {\bibfnamefont
  {W.~D.}\ \bibnamefont {Richards}}, \bibinfo {author} {\bibfnamefont {A.~C.}\
  \bibnamefont {Gamst}}, \bibinfo {author} {\bibfnamefont {K.~A.}\ \bibnamefont
  {Persson}}, \ and\ \bibinfo {author} {\bibfnamefont {G.}~\bibnamefont
  {Ceder}},\ }\href {\doibase 10.1126/sciadv.1600225} {\bibfield  {journal}
  {\bibinfo  {journal} {Science Advances}\ }\textbf {\bibinfo {volume} {2}}
  (\bibinfo {year} {2016}{\natexlab{b}}),\ 10.1126/sciadv.1600225},\ \Eprint
  {http://arxiv.org/abs/http://advances.sciencemag.org/content/2/11/e1600225.full.pdf}
  {http://advances.sciencemag.org/content/2/11/e1600225.full.pdf} \BibitemShut
  {NoStop}%
\bibitem [{\citenamefont {Amsler}\ \emph {et~al.}(2018)\citenamefont {Amsler},
  \citenamefont {Hegde}, \citenamefont {Jacobsen},\ and\ \citenamefont
  {Wolverton}}]{amsler2018exploring}%
  \BibitemOpen
  \bibfield  {author} {\bibinfo {author} {\bibfnamefont {M.}~\bibnamefont
  {Amsler}}, \bibinfo {author} {\bibfnamefont {V.~I.}\ \bibnamefont {Hegde}},
  \bibinfo {author} {\bibfnamefont {S.~D.}\ \bibnamefont {Jacobsen}}, \ and\
  \bibinfo {author} {\bibfnamefont {C.}~\bibnamefont {Wolverton}},\ }\href@noop
  {} {\bibfield  {journal} {\bibinfo  {journal} {arXiv preprint
  arXiv:1802.06900}\ } (\bibinfo {year} {2018})}\BibitemShut {NoStop}%
\bibitem [{\citenamefont {Jagodzinski}(1949)}]{Jagodzinski:a00148}%
  \BibitemOpen
  \bibfield  {author} {\bibinfo {author} {\bibfnamefont {H.}~\bibnamefont
  {Jagodzinski}},\ }\href {\doibase 10.1107/S0365110X49000552} {\bibfield
  {journal} {\bibinfo  {journal} {Acta Crystallographica}\ }\textbf {\bibinfo
  {volume} {2}},\ \bibinfo {pages} {201} (\bibinfo {year} {1949})}\BibitemShut
  {NoStop}%
\bibitem [{\citenamefont {Tilley}(2016)}]{tilley2016perovskites}%
  \BibitemOpen
  \bibfield  {author} {\bibinfo {author} {\bibfnamefont {R.}~\bibnamefont
  {Tilley}},\ }\href {https://books.google.ch/books?id=jDS7CwAAQBAJ} {\emph
  {\bibinfo {title} {Perovskites: Structure-Property Relationships}}}\
  (\bibinfo  {publisher} {Wiley},\ \bibinfo {year} {2016})\BibitemShut
  {NoStop}%
\bibitem [{\citenamefont {Swainson}\ \emph {et~al.}(2007)\citenamefont
  {Swainson}, \citenamefont {Tucker}, \citenamefont {Wilson}, \citenamefont
  {Winkler},\ and\ \citenamefont {Milman}}]{swainson2007pressure}%
  \BibitemOpen
  \bibfield  {author} {\bibinfo {author} {\bibfnamefont {I.}~\bibnamefont
  {Swainson}}, \bibinfo {author} {\bibfnamefont {M.}~\bibnamefont {Tucker}},
  \bibinfo {author} {\bibfnamefont {D.}~\bibnamefont {Wilson}}, \bibinfo
  {author} {\bibfnamefont {B.}~\bibnamefont {Winkler}}, \ and\ \bibinfo
  {author} {\bibfnamefont {V.}~\bibnamefont {Milman}},\ }\href@noop {}
  {\bibfield  {journal} {\bibinfo  {journal} {Chemistry of materials}\ }\textbf
  {\bibinfo {volume} {19}},\ \bibinfo {pages} {2401} (\bibinfo {year}
  {2007})}\BibitemShut {NoStop}%
\bibitem [{\citenamefont {Wang}\ \emph {et~al.}(2015)\citenamefont {Wang},
  \citenamefont {L{\"u}}, \citenamefont {Yang}, \citenamefont {Wen},
  \citenamefont {Yang}, \citenamefont {Ren}, \citenamefont {Wang},
  \citenamefont {Lin},\ and\ \citenamefont {Zhao}}]{wang2015pressure}%
  \BibitemOpen
  \bibfield  {author} {\bibinfo {author} {\bibfnamefont {Y.}~\bibnamefont
  {Wang}}, \bibinfo {author} {\bibfnamefont {X.}~\bibnamefont {L{\"u}}},
  \bibinfo {author} {\bibfnamefont {W.}~\bibnamefont {Yang}}, \bibinfo {author}
  {\bibfnamefont {T.}~\bibnamefont {Wen}}, \bibinfo {author} {\bibfnamefont
  {L.}~\bibnamefont {Yang}}, \bibinfo {author} {\bibfnamefont {X.}~\bibnamefont
  {Ren}}, \bibinfo {author} {\bibfnamefont {L.}~\bibnamefont {Wang}}, \bibinfo
  {author} {\bibfnamefont {Z.}~\bibnamefont {Lin}}, \ and\ \bibinfo {author}
  {\bibfnamefont {Y.}~\bibnamefont {Zhao}},\ }\href@noop {} {\bibfield
  {journal} {\bibinfo  {journal} {Journal of the American Chemical Society}\
  }\textbf {\bibinfo {volume} {137}},\ \bibinfo {pages} {11144} (\bibinfo
  {year} {2015})}\BibitemShut {NoStop}%
\bibitem [{\citenamefont {Postorino}\ and\ \citenamefont
  {Malavasi}(2017)}]{postorino2017pressure}%
  \BibitemOpen
  \bibfield  {author} {\bibinfo {author} {\bibfnamefont {P.}~\bibnamefont
  {Postorino}}\ and\ \bibinfo {author} {\bibfnamefont {L.}~\bibnamefont
  {Malavasi}},\ }\href@noop {} {\bibfield  {journal} {\bibinfo  {journal} {The
  Journal of Physical Chemistry Letters}\ }\textbf {\bibinfo {volume} {8}},\
  \bibinfo {pages} {2613} (\bibinfo {year} {2017})}\BibitemShut {NoStop}%
\bibitem [{\citenamefont {Umari}\ \emph {et~al.}(2014)\citenamefont {Umari},
  \citenamefont {Mosconi},\ and\ \citenamefont
  {De~Angelis}}]{umari_GWSOC_2014}%
  \BibitemOpen
  \bibfield  {author} {\bibinfo {author} {\bibfnamefont {P.}~\bibnamefont
  {Umari}}, \bibinfo {author} {\bibfnamefont {E.}~\bibnamefont {Mosconi}}, \
  and\ \bibinfo {author} {\bibfnamefont {F.}~\bibnamefont {De~Angelis}},\
  }\href@noop {} {\bibfield  {journal} {\bibinfo  {journal} {Scientific
  reports}\ }\textbf {\bibinfo {volume} {4}},\ \bibinfo {pages} {4467}
  (\bibinfo {year} {2014})}\BibitemShut {NoStop}%
\bibitem [{\citenamefont {Blancon}\ \emph {et~al.}(2017)\citenamefont
  {Blancon}, \citenamefont {Tsai}, \citenamefont {Nie}, \citenamefont
  {Stoumpos}, \citenamefont {Pedesseau}, \citenamefont {Katan}, \citenamefont
  {Kepenekian}, \citenamefont {Soe}, \citenamefont {Appavoo}, \citenamefont
  {Sfeir}, \citenamefont {Tretiak}, \citenamefont {Ajayan}, \citenamefont
  {Kanatzidis}, \citenamefont {Even}, \citenamefont {Crochet},\ and\
  \citenamefont {Mohite}}]{blancon2017extremely}%
  \BibitemOpen
  \bibfield  {author} {\bibinfo {author} {\bibfnamefont {J.-C.}\ \bibnamefont
  {Blancon}}, \bibinfo {author} {\bibfnamefont {H.}~\bibnamefont {Tsai}},
  \bibinfo {author} {\bibfnamefont {W.}~\bibnamefont {Nie}}, \bibinfo {author}
  {\bibfnamefont {C.~C.}\ \bibnamefont {Stoumpos}}, \bibinfo {author}
  {\bibfnamefont {L.}~\bibnamefont {Pedesseau}}, \bibinfo {author}
  {\bibfnamefont {C.}~\bibnamefont {Katan}}, \bibinfo {author} {\bibfnamefont
  {M.}~\bibnamefont {Kepenekian}}, \bibinfo {author} {\bibfnamefont {C.~M.~M.}\
  \bibnamefont {Soe}}, \bibinfo {author} {\bibfnamefont {K.}~\bibnamefont
  {Appavoo}}, \bibinfo {author} {\bibfnamefont {M.~Y.}\ \bibnamefont {Sfeir}},
  \bibinfo {author} {\bibfnamefont {S.}~\bibnamefont {Tretiak}}, \bibinfo
  {author} {\bibfnamefont {P.~M.}\ \bibnamefont {Ajayan}}, \bibinfo {author}
  {\bibfnamefont {M.~G.}\ \bibnamefont {Kanatzidis}}, \bibinfo {author}
  {\bibfnamefont {J.}~\bibnamefont {Even}}, \bibinfo {author} {\bibfnamefont
  {J.~J.}\ \bibnamefont {Crochet}}, \ and\ \bibinfo {author} {\bibfnamefont
  {A.~D.}\ \bibnamefont {Mohite}},\ }\href {\doibase 10.1126/science.aal4211}
  {\bibfield  {journal} {\bibinfo  {journal} {Science}\ }\textbf {\bibinfo
  {volume} {355}},\ \bibinfo {pages} {1288} (\bibinfo {year} {2017})},\ \Eprint
  {http://arxiv.org/abs/http://science.sciencemag.org/content/355/6331/1288.full.pdf}
  {http://science.sciencemag.org/content/355/6331/1288.full.pdf} \BibitemShut
  {NoStop}%
\bibitem [{\citenamefont {Dar}\ \emph {et~al.}(2016)\citenamefont {Dar},
  \citenamefont {Jacopin}, \citenamefont {Meloni}, \citenamefont {Mattoni},
  \citenamefont {Arora}, \citenamefont {Boziki}, \citenamefont {Zakeeruddin},
  \citenamefont {Rothlisberger},\ and\ \citenamefont
  {Gr{\"a}tzel}}]{dar2016origin}%
  \BibitemOpen
  \bibfield  {author} {\bibinfo {author} {\bibfnamefont {M.~I.}\ \bibnamefont
  {Dar}}, \bibinfo {author} {\bibfnamefont {G.}~\bibnamefont {Jacopin}},
  \bibinfo {author} {\bibfnamefont {S.}~\bibnamefont {Meloni}}, \bibinfo
  {author} {\bibfnamefont {A.}~\bibnamefont {Mattoni}}, \bibinfo {author}
  {\bibfnamefont {N.}~\bibnamefont {Arora}}, \bibinfo {author} {\bibfnamefont
  {A.}~\bibnamefont {Boziki}}, \bibinfo {author} {\bibfnamefont {S.~M.}\
  \bibnamefont {Zakeeruddin}}, \bibinfo {author} {\bibfnamefont
  {U.}~\bibnamefont {Rothlisberger}}, \ and\ \bibinfo {author} {\bibfnamefont
  {M.}~\bibnamefont {Gr{\"a}tzel}},\ }\href@noop {} {\bibfield  {journal}
  {\bibinfo  {journal} {Science advances}\ }\textbf {\bibinfo {volume} {2}},\
  \bibinfo {pages} {e1601156} (\bibinfo {year} {2016})}\BibitemShut {NoStop}%
\bibitem [{\citenamefont {Eperon}\ \emph {et~al.}(2014)\citenamefont {Eperon},
  \citenamefont {Stranks}, \citenamefont {Menelaou}, \citenamefont {Johnston},
  \citenamefont {Herz},\ and\ \citenamefont
  {Snaith}}]{eperon2014formamidinium}%
  \BibitemOpen
  \bibfield  {author} {\bibinfo {author} {\bibfnamefont {G.~E.}\ \bibnamefont
  {Eperon}}, \bibinfo {author} {\bibfnamefont {S.~D.}\ \bibnamefont {Stranks}},
  \bibinfo {author} {\bibfnamefont {C.}~\bibnamefont {Menelaou}}, \bibinfo
  {author} {\bibfnamefont {M.~B.}\ \bibnamefont {Johnston}}, \bibinfo {author}
  {\bibfnamefont {L.~M.}\ \bibnamefont {Herz}}, \ and\ \bibinfo {author}
  {\bibfnamefont {H.~J.}\ \bibnamefont {Snaith}},\ }\href@noop {} {\bibfield
  {journal} {\bibinfo  {journal} {Energy \& Environmental Science}\ }\textbf
  {\bibinfo {volume} {7}},\ \bibinfo {pages} {982} (\bibinfo {year}
  {2014})}\BibitemShut {NoStop}%
\bibitem [{\citenamefont {Baikie}\ \emph {et~al.}(2013)\citenamefont {Baikie},
  \citenamefont {Fang}, \citenamefont {Kadro}, \citenamefont {Schreyer},
  \citenamefont {Wei}, \citenamefont {Mhaisalkar}, \citenamefont {Graetzel},\
  and\ \citenamefont {White}}]{baikie2013synthesis}%
  \BibitemOpen
  \bibfield  {author} {\bibinfo {author} {\bibfnamefont {T.}~\bibnamefont
  {Baikie}}, \bibinfo {author} {\bibfnamefont {Y.}~\bibnamefont {Fang}},
  \bibinfo {author} {\bibfnamefont {J.~M.}\ \bibnamefont {Kadro}}, \bibinfo
  {author} {\bibfnamefont {M.}~\bibnamefont {Schreyer}}, \bibinfo {author}
  {\bibfnamefont {F.}~\bibnamefont {Wei}}, \bibinfo {author} {\bibfnamefont
  {S.~G.}\ \bibnamefont {Mhaisalkar}}, \bibinfo {author} {\bibfnamefont
  {M.}~\bibnamefont {Graetzel}}, \ and\ \bibinfo {author} {\bibfnamefont
  {T.~J.}\ \bibnamefont {White}},\ }\href@noop {} {\bibfield  {journal}
  {\bibinfo  {journal} {Journal of Materials Chemistry A}\ }\textbf {\bibinfo
  {volume} {1}},\ \bibinfo {pages} {5628} (\bibinfo {year} {2013})}\BibitemShut
  {NoStop}%
\bibitem [{\citenamefont {Milot}\ \emph {et~al.}(2015)\citenamefont {Milot},
  \citenamefont {Eperon}, \citenamefont {Snaith}, \citenamefont {Johnston},\
  and\ \citenamefont {Herz}}]{milot2015temperature}%
  \BibitemOpen
  \bibfield  {author} {\bibinfo {author} {\bibfnamefont {R.~L.}\ \bibnamefont
  {Milot}}, \bibinfo {author} {\bibfnamefont {G.~E.}\ \bibnamefont {Eperon}},
  \bibinfo {author} {\bibfnamefont {H.~J.}\ \bibnamefont {Snaith}}, \bibinfo
  {author} {\bibfnamefont {M.~B.}\ \bibnamefont {Johnston}}, \ and\ \bibinfo
  {author} {\bibfnamefont {L.~M.}\ \bibnamefont {Herz}},\ }\href@noop {}
  {\bibfield  {journal} {\bibinfo  {journal} {Advanced Functional Materials}\
  }\textbf {\bibinfo {volume} {25}},\ \bibinfo {pages} {6218} (\bibinfo {year}
  {2015})}\BibitemShut {NoStop}%
\bibitem [{\citenamefont {Motta}\ \emph {et~al.}(2015)\citenamefont {Motta},
  \citenamefont {El-Mellouhi}, \citenamefont {Kais}, \citenamefont {Tabet},
  \citenamefont {Alharbi},\ and\ \citenamefont {Sanvito}}]{motta2015revealing}%
  \BibitemOpen
  \bibfield  {author} {\bibinfo {author} {\bibfnamefont {C.}~\bibnamefont
  {Motta}}, \bibinfo {author} {\bibfnamefont {F.}~\bibnamefont {El-Mellouhi}},
  \bibinfo {author} {\bibfnamefont {S.}~\bibnamefont {Kais}}, \bibinfo {author}
  {\bibfnamefont {N.}~\bibnamefont {Tabet}}, \bibinfo {author} {\bibfnamefont
  {F.}~\bibnamefont {Alharbi}}, \ and\ \bibinfo {author} {\bibfnamefont
  {S.}~\bibnamefont {Sanvito}},\ }\href@noop {} {\bibfield  {journal} {\bibinfo
   {journal} {Nature communications}\ }\textbf {\bibinfo {volume} {6}},\
  \bibinfo {pages} {7026} (\bibinfo {year} {2015})}\BibitemShut {NoStop}%
\bibitem [{\citenamefont {Leguy}\ \emph {et~al.}(2016)\citenamefont {Leguy},
  \citenamefont {Azarhoosh}, \citenamefont {Alonso}, \citenamefont
  {Campoy-Quiles}, \citenamefont {Weber}, \citenamefont {Yao}, \citenamefont
  {Bryant}, \citenamefont {Weller}, \citenamefont {Nelson}, \citenamefont
  {Walsh}, \citenamefont {van Schilfgaarde},\ and\ \citenamefont
  {Barnes}}]{leguy2016experimental}%
  \BibitemOpen
  \bibfield  {author} {\bibinfo {author} {\bibfnamefont {A.~M.~A.}\
  \bibnamefont {Leguy}}, \bibinfo {author} {\bibfnamefont {P.}~\bibnamefont
  {Azarhoosh}}, \bibinfo {author} {\bibfnamefont {M.~I.}\ \bibnamefont
  {Alonso}}, \bibinfo {author} {\bibfnamefont {M.}~\bibnamefont
  {Campoy-Quiles}}, \bibinfo {author} {\bibfnamefont {O.~J.}\ \bibnamefont
  {Weber}}, \bibinfo {author} {\bibfnamefont {J.}~\bibnamefont {Yao}}, \bibinfo
  {author} {\bibfnamefont {D.}~\bibnamefont {Bryant}}, \bibinfo {author}
  {\bibfnamefont {M.~T.}\ \bibnamefont {Weller}}, \bibinfo {author}
  {\bibfnamefont {J.}~\bibnamefont {Nelson}}, \bibinfo {author} {\bibfnamefont
  {A.}~\bibnamefont {Walsh}}, \bibinfo {author} {\bibfnamefont
  {M.}~\bibnamefont {van Schilfgaarde}}, \ and\ \bibinfo {author}
  {\bibfnamefont {P.~R.~F.}\ \bibnamefont {Barnes}},\ }\href {\doibase
  10.1039/C5NR05435D} {\bibfield  {journal} {\bibinfo  {journal} {Nanoscale}\
  }\textbf {\bibinfo {volume} {8}},\ \bibinfo {pages} {6317} (\bibinfo {year}
  {2016})}\BibitemShut {NoStop}%
\bibitem [{\citenamefont {Kim}\ \emph {et~al.}(2017{\natexlab{b}})\citenamefont
  {Kim}, \citenamefont {Hunger}, \citenamefont {C{\'a}novas}, \citenamefont
  {Karakus}, \citenamefont {Mics}, \citenamefont {Grechko}, \citenamefont
  {Turchinovich}, \citenamefont {Parekh},\ and\ \citenamefont
  {Bonn}}]{kim2017direct}%
  \BibitemOpen
  \bibfield  {author} {\bibinfo {author} {\bibfnamefont {H.}~\bibnamefont
  {Kim}}, \bibinfo {author} {\bibfnamefont {J.}~\bibnamefont {Hunger}},
  \bibinfo {author} {\bibfnamefont {E.}~\bibnamefont {C{\'a}novas}}, \bibinfo
  {author} {\bibfnamefont {M.}~\bibnamefont {Karakus}}, \bibinfo {author}
  {\bibfnamefont {Z.}~\bibnamefont {Mics}}, \bibinfo {author} {\bibfnamefont
  {M.}~\bibnamefont {Grechko}}, \bibinfo {author} {\bibfnamefont
  {D.}~\bibnamefont {Turchinovich}}, \bibinfo {author} {\bibfnamefont {S.~H.}\
  \bibnamefont {Parekh}}, \ and\ \bibinfo {author} {\bibfnamefont
  {M.}~\bibnamefont {Bonn}},\ }\href@noop {} {\bibfield  {journal} {\bibinfo
  {journal} {Nature communications}\ }\textbf {\bibinfo {volume} {8}},\
  \bibinfo {pages} {687} (\bibinfo {year} {2017}{\natexlab{b}})}\BibitemShut
  {NoStop}%
\bibitem [{\citenamefont {Toledano}\ and\ \citenamefont
  {Toledano}(1987)}]{toledano1987landau}%
  \BibitemOpen
  \bibfield  {author} {\bibinfo {author} {\bibfnamefont {J.-C.}\ \bibnamefont
  {Toledano}}\ and\ \bibinfo {author} {\bibfnamefont {P.}~\bibnamefont
  {Toledano}},\ }\href@noop {} {\emph {\bibinfo {title} {The Landau theory of
  phase transitions: application to structural, incommensurate, magnetic and
  liquid crystal systems}}},\ Vol.~\bibinfo {volume} {3}\ (\bibinfo
  {publisher} {World Scientific Publishing Company},\ \bibinfo {year}
  {1987})\BibitemShut {NoStop}%
\bibitem [{\citenamefont {Howard}\ and\ \citenamefont
  {Stokes}(1998)}]{howard1998group}%
  \BibitemOpen
  \bibfield  {author} {\bibinfo {author} {\bibfnamefont {C.~J.}\ \bibnamefont
  {Howard}}\ and\ \bibinfo {author} {\bibfnamefont {H.~T.}\ \bibnamefont
  {Stokes}},\ }\href@noop {} {\bibfield  {journal} {\bibinfo  {journal} {Acta
  Crystallographica Section B: Structural Science}\ }\textbf {\bibinfo {volume}
  {54}},\ \bibinfo {pages} {782} (\bibinfo {year} {1998})}\BibitemShut
  {NoStop}%
\bibitem [{\citenamefont {Brivio}\ \emph {et~al.}(2015)\citenamefont {Brivio},
  \citenamefont {Frost}, \citenamefont {Skelton}, \citenamefont {Jackson},
  \citenamefont {Weber}, \citenamefont {Weller}, \citenamefont {Go\~ni},
  \citenamefont {Leguy}, \citenamefont {Barnes},\ and\ \citenamefont
  {Walsh}}]{Brivio_Frost_PRB_2015}%
  \BibitemOpen
  \bibfield  {author} {\bibinfo {author} {\bibfnamefont {F.}~\bibnamefont
  {Brivio}}, \bibinfo {author} {\bibfnamefont {J.~M.}\ \bibnamefont {Frost}},
  \bibinfo {author} {\bibfnamefont {J.~M.}\ \bibnamefont {Skelton}}, \bibinfo
  {author} {\bibfnamefont {A.~J.}\ \bibnamefont {Jackson}}, \bibinfo {author}
  {\bibfnamefont {O.~J.}\ \bibnamefont {Weber}}, \bibinfo {author}
  {\bibfnamefont {M.~T.}\ \bibnamefont {Weller}}, \bibinfo {author}
  {\bibfnamefont {A.~R.}\ \bibnamefont {Go\~ni}}, \bibinfo {author}
  {\bibfnamefont {A.~M.~A.}\ \bibnamefont {Leguy}}, \bibinfo {author}
  {\bibfnamefont {P.~R.~F.}\ \bibnamefont {Barnes}}, \ and\ \bibinfo {author}
  {\bibfnamefont {A.}~\bibnamefont {Walsh}},\ }\href {\doibase
  10.1103/PhysRevB.92.144308} {\bibfield  {journal} {\bibinfo  {journal} {Phys.
  Rev. B}\ }\textbf {\bibinfo {volume} {92}},\ \bibinfo {pages} {144308}
  (\bibinfo {year} {2015})}\BibitemShut {NoStop}%
\bibitem [{\citenamefont {Kieslich}\ and\ \citenamefont
  {Goodwin}(2017)}]{thesame_not_thesame}%
  \BibitemOpen
  \bibfield  {author} {\bibinfo {author} {\bibfnamefont {G.}~\bibnamefont
  {Kieslich}}\ and\ \bibinfo {author} {\bibfnamefont {A.~L.}\ \bibnamefont
  {Goodwin}},\ }\href@noop {} {\bibfield  {journal} {\bibinfo  {journal}
  {Materials Horizons}\ }\textbf {\bibinfo {volume} {4}},\ \bibinfo {pages}
  {362} (\bibinfo {year} {2017})}\BibitemShut {NoStop}%
\bibitem [{\citenamefont {Yi}\ \emph {et~al.}(2016)\citenamefont {Yi},
  \citenamefont {Luo}, \citenamefont {Meloni}, \citenamefont {Boziki},
  \citenamefont {Ashari-Astani}, \citenamefont {Gr{\"a}tzel}, \citenamefont
  {Zakeeruddin}, \citenamefont {R{\"o}thlisberger},\ and\ \citenamefont
  {Gr{\"a}tzel}}]{yi2016entropic}%
  \BibitemOpen
  \bibfield  {author} {\bibinfo {author} {\bibfnamefont {C.}~\bibnamefont
  {Yi}}, \bibinfo {author} {\bibfnamefont {J.}~\bibnamefont {Luo}}, \bibinfo
  {author} {\bibfnamefont {S.}~\bibnamefont {Meloni}}, \bibinfo {author}
  {\bibfnamefont {A.}~\bibnamefont {Boziki}}, \bibinfo {author} {\bibfnamefont
  {N.}~\bibnamefont {Ashari-Astani}}, \bibinfo {author} {\bibfnamefont
  {C.}~\bibnamefont {Gr{\"a}tzel}}, \bibinfo {author} {\bibfnamefont {S.~M.}\
  \bibnamefont {Zakeeruddin}}, \bibinfo {author} {\bibfnamefont
  {U.}~\bibnamefont {R{\"o}thlisberger}}, \ and\ \bibinfo {author}
  {\bibfnamefont {M.}~\bibnamefont {Gr{\"a}tzel}},\ }\href@noop {} {\bibfield
  {journal} {\bibinfo  {journal} {Energy \& Environmental Science}\ }\textbf
  {\bibinfo {volume} {9}},\ \bibinfo {pages} {656} (\bibinfo {year}
  {2016})}\BibitemShut {NoStop}%
\bibitem [{\citenamefont {Ma}\ \emph {et~al.}(2017)\citenamefont {Ma},
  \citenamefont {Li}, \citenamefont {Li}, \citenamefont {Lin}, \citenamefont
  {Wang},\ and\ \citenamefont {Qiao}}]{ma2017stable}%
  \BibitemOpen
  \bibfield  {author} {\bibinfo {author} {\bibfnamefont {F.}~\bibnamefont
  {Ma}}, \bibinfo {author} {\bibfnamefont {J.}~\bibnamefont {Li}}, \bibinfo
  {author} {\bibfnamefont {W.}~\bibnamefont {Li}}, \bibinfo {author}
  {\bibfnamefont {N.}~\bibnamefont {Lin}}, \bibinfo {author} {\bibfnamefont
  {L.}~\bibnamefont {Wang}}, \ and\ \bibinfo {author} {\bibfnamefont
  {J.}~\bibnamefont {Qiao}},\ }\href@noop {} {\bibfield  {journal} {\bibinfo
  {journal} {Chemical science}\ }\textbf {\bibinfo {volume} {8}},\ \bibinfo
  {pages} {800} (\bibinfo {year} {2017})}\BibitemShut {NoStop}%
\bibitem [{\citenamefont {Eperon}\ \emph {et~al.}(2015)\citenamefont {Eperon},
  \citenamefont {Paterno}, \citenamefont {Sutton}, \citenamefont {Zampetti},
  \citenamefont {Haghighirad}, \citenamefont {Cacialli},\ and\ \citenamefont
  {Snaith}}]{eperon2015inorganic}%
  \BibitemOpen
  \bibfield  {author} {\bibinfo {author} {\bibfnamefont {G.~E.}\ \bibnamefont
  {Eperon}}, \bibinfo {author} {\bibfnamefont {G.~M.}\ \bibnamefont {Paterno}},
  \bibinfo {author} {\bibfnamefont {R.~J.}\ \bibnamefont {Sutton}}, \bibinfo
  {author} {\bibfnamefont {A.}~\bibnamefont {Zampetti}}, \bibinfo {author}
  {\bibfnamefont {A.~A.}\ \bibnamefont {Haghighirad}}, \bibinfo {author}
  {\bibfnamefont {F.}~\bibnamefont {Cacialli}}, \ and\ \bibinfo {author}
  {\bibfnamefont {H.~J.}\ \bibnamefont {Snaith}},\ }\href@noop {} {\bibfield
  {journal} {\bibinfo  {journal} {Journal of Materials Chemistry A}\ }\textbf
  {\bibinfo {volume} {3}},\ \bibinfo {pages} {19688} (\bibinfo {year}
  {2015})}\BibitemShut {NoStop}%
\bibitem [{\citenamefont {Lai}\ \emph {et~al.}(2017)\citenamefont {Lai},
  \citenamefont {Kong}, \citenamefont {Bischak}, \citenamefont {Yu},
  \citenamefont {Dou}, \citenamefont {Eaton}, \citenamefont {Ginsberg},\ and\
  \citenamefont {Yang}}]{lai2017structural}%
  \BibitemOpen
  \bibfield  {author} {\bibinfo {author} {\bibfnamefont {M.}~\bibnamefont
  {Lai}}, \bibinfo {author} {\bibfnamefont {Q.}~\bibnamefont {Kong}}, \bibinfo
  {author} {\bibfnamefont {C.~G.}\ \bibnamefont {Bischak}}, \bibinfo {author}
  {\bibfnamefont {Y.}~\bibnamefont {Yu}}, \bibinfo {author} {\bibfnamefont
  {L.}~\bibnamefont {Dou}}, \bibinfo {author} {\bibfnamefont {S.~W.}\
  \bibnamefont {Eaton}}, \bibinfo {author} {\bibfnamefont {N.~S.}\ \bibnamefont
  {Ginsberg}}, \ and\ \bibinfo {author} {\bibfnamefont {P.}~\bibnamefont
  {Yang}},\ }\href@noop {} {\bibfield  {journal} {\bibinfo  {journal} {Nano
  Research}\ }\textbf {\bibinfo {volume} {10}},\ \bibinfo {pages} {1107}
  (\bibinfo {year} {2017})}\BibitemShut {NoStop}%
\bibitem [{\citenamefont {Stoumpos}\ \emph {et~al.}(2016)\citenamefont
  {Stoumpos}, \citenamefont {Mao}, \citenamefont {Malliakas},\ and\
  \citenamefont {Kanatzidis}}]{stoumpos2016structure}%
  \BibitemOpen
  \bibfield  {author} {\bibinfo {author} {\bibfnamefont {C.~C.}\ \bibnamefont
  {Stoumpos}}, \bibinfo {author} {\bibfnamefont {L.}~\bibnamefont {Mao}},
  \bibinfo {author} {\bibfnamefont {C.~D.}\ \bibnamefont {Malliakas}}, \ and\
  \bibinfo {author} {\bibfnamefont {M.~G.}\ \bibnamefont {Kanatzidis}},\
  }\href@noop {} {\bibfield  {journal} {\bibinfo  {journal} {Inorganic
  chemistry}\ }\textbf {\bibinfo {volume} {56}},\ \bibinfo {pages} {56}
  (\bibinfo {year} {2016})}\BibitemShut {NoStop}%
\bibitem [{\citenamefont {Manser}\ \emph {et~al.}(2016)\citenamefont {Manser},
  \citenamefont {Saidaminov}, \citenamefont {Christians}, \citenamefont
  {Bakr},\ and\ \citenamefont {Kamat}}]{manser2016making}%
  \BibitemOpen
  \bibfield  {author} {\bibinfo {author} {\bibfnamefont {J.~S.}\ \bibnamefont
  {Manser}}, \bibinfo {author} {\bibfnamefont {M.~I.}\ \bibnamefont
  {Saidaminov}}, \bibinfo {author} {\bibfnamefont {J.~A.}\ \bibnamefont
  {Christians}}, \bibinfo {author} {\bibfnamefont {O.~M.}\ \bibnamefont
  {Bakr}}, \ and\ \bibinfo {author} {\bibfnamefont {P.~V.}\ \bibnamefont
  {Kamat}},\ }\href@noop {} {\bibfield  {journal} {\bibinfo  {journal}
  {Accounts of chemical research}\ }\textbf {\bibinfo {volume} {49}},\ \bibinfo
  {pages} {330} (\bibinfo {year} {2016})}\BibitemShut {NoStop}%
\bibitem [{\citenamefont {Frost}\ \emph {et~al.}(2014)\citenamefont {Frost},
  \citenamefont {Butler}, \citenamefont {Brivio}, \citenamefont {Hendon},
  \citenamefont {Van~Schilfgaarde},\ and\ \citenamefont
  {Walsh}}]{frost2014atomistic}%
  \BibitemOpen
  \bibfield  {author} {\bibinfo {author} {\bibfnamefont {J.~M.}\ \bibnamefont
  {Frost}}, \bibinfo {author} {\bibfnamefont {K.~T.}\ \bibnamefont {Butler}},
  \bibinfo {author} {\bibfnamefont {F.}~\bibnamefont {Brivio}}, \bibinfo
  {author} {\bibfnamefont {C.~H.}\ \bibnamefont {Hendon}}, \bibinfo {author}
  {\bibfnamefont {M.}~\bibnamefont {Van~Schilfgaarde}}, \ and\ \bibinfo
  {author} {\bibfnamefont {A.}~\bibnamefont {Walsh}},\ }\href@noop {}
  {\bibfield  {journal} {\bibinfo  {journal} {Nano letters}\ }\textbf {\bibinfo
  {volume} {14}},\ \bibinfo {pages} {2584} (\bibinfo {year}
  {2014})}\BibitemShut {NoStop}%
\bibitem [{\citenamefont {Jaffe}\ \emph
  {et~al.}(2017{\natexlab{b}})\citenamefont {Jaffe}, \citenamefont {Lin},
  \citenamefont {Mao},\ and\ \citenamefont {Karunadasa}}]{jaffe2017pressure}%
  \BibitemOpen
  \bibfield  {author} {\bibinfo {author} {\bibfnamefont {A.}~\bibnamefont
  {Jaffe}}, \bibinfo {author} {\bibfnamefont {Y.}~\bibnamefont {Lin}}, \bibinfo
  {author} {\bibfnamefont {W.~L.}\ \bibnamefont {Mao}}, \ and\ \bibinfo
  {author} {\bibfnamefont {H.~I.}\ \bibnamefont {Karunadasa}},\ }\href@noop {}
  {\bibfield  {journal} {\bibinfo  {journal} {Journal of the American Chemical
  Society}\ }\textbf {\bibinfo {volume} {139}},\ \bibinfo {pages} {4330}
  (\bibinfo {year} {2017}{\natexlab{b}})}\BibitemShut {NoStop}%
\bibitem [{\citenamefont {Mao}\ \emph {et~al.}(2003)\citenamefont {Mao},
  \citenamefont {Mao}, \citenamefont {Eng}, \citenamefont {Trainor},
  \citenamefont {Newville}, \citenamefont {Kao}, \citenamefont {Heinz},
  \citenamefont {Shu}, \citenamefont {Meng},\ and\ \citenamefont
  {Hemley}}]{mao2003bonding}%
  \BibitemOpen
  \bibfield  {author} {\bibinfo {author} {\bibfnamefont {W.~L.}\ \bibnamefont
  {Mao}}, \bibinfo {author} {\bibfnamefont {H.-k.}\ \bibnamefont {Mao}},
  \bibinfo {author} {\bibfnamefont {P.~J.}\ \bibnamefont {Eng}}, \bibinfo
  {author} {\bibfnamefont {T.~P.}\ \bibnamefont {Trainor}}, \bibinfo {author}
  {\bibfnamefont {M.}~\bibnamefont {Newville}}, \bibinfo {author}
  {\bibfnamefont {C.-c.}\ \bibnamefont {Kao}}, \bibinfo {author} {\bibfnamefont
  {D.~L.}\ \bibnamefont {Heinz}}, \bibinfo {author} {\bibfnamefont
  {J.}~\bibnamefont {Shu}}, \bibinfo {author} {\bibfnamefont {Y.}~\bibnamefont
  {Meng}}, \ and\ \bibinfo {author} {\bibfnamefont {R.~J.}\ \bibnamefont
  {Hemley}},\ }\href@noop {} {\bibfield  {journal} {\bibinfo  {journal}
  {Science}\ }\textbf {\bibinfo {volume} {302}},\ \bibinfo {pages} {425}
  (\bibinfo {year} {2003})}\BibitemShut {NoStop}%
\bibitem [{\citenamefont {Amsler}\ \emph {et~al.}(2012)\citenamefont {Amsler},
  \citenamefont {Flores-Livas}, \citenamefont {Lehtovaara}, \citenamefont
  {Balima}, \citenamefont {Ghasemi}, \citenamefont {Machon}, \citenamefont
  {Pailh\`es}, \citenamefont {Willand}, \citenamefont {Caliste}, \citenamefont
  {Botti}, \citenamefont {San~Miguel}, \citenamefont {Goedecker},\ and\
  \citenamefont {Marques}}]{maxi_2012-Zcarbon}%
  \BibitemOpen
  \bibfield  {author} {\bibinfo {author} {\bibfnamefont {M.}~\bibnamefont
  {Amsler}}, \bibinfo {author} {\bibfnamefont {J.~A.}\ \bibnamefont
  {Flores-Livas}}, \bibinfo {author} {\bibfnamefont {L.}~\bibnamefont
  {Lehtovaara}}, \bibinfo {author} {\bibfnamefont {F.}~\bibnamefont {Balima}},
  \bibinfo {author} {\bibfnamefont {S.~A.}\ \bibnamefont {Ghasemi}}, \bibinfo
  {author} {\bibfnamefont {D.}~\bibnamefont {Machon}}, \bibinfo {author}
  {\bibfnamefont {S.}~\bibnamefont {Pailh\`es}}, \bibinfo {author}
  {\bibfnamefont {A.}~\bibnamefont {Willand}}, \bibinfo {author} {\bibfnamefont
  {D.}~\bibnamefont {Caliste}}, \bibinfo {author} {\bibfnamefont
  {S.}~\bibnamefont {Botti}}, \bibinfo {author} {\bibfnamefont
  {A.}~\bibnamefont {San~Miguel}}, \bibinfo {author} {\bibfnamefont
  {S.}~\bibnamefont {Goedecker}}, \ and\ \bibinfo {author} {\bibfnamefont
  {M.~A.~L.}\ \bibnamefont {Marques}},\ }\href {\doibase
  10.1103/PhysRevLett.108.065501} {\bibfield  {journal} {\bibinfo  {journal}
  {Phys. Rev. Lett.}\ }\textbf {\bibinfo {volume} {108}},\ \bibinfo {pages}
  {065501} (\bibinfo {year} {2012})}\BibitemShut {NoStop}%
\bibitem [{\citenamefont {Flores-Livas}\ \emph
  {et~al.}(2017{\natexlab{b}})\citenamefont {Flores-Livas}, \citenamefont
  {Sanna}, \citenamefont {Drozdov}, \citenamefont {Boeri}, \citenamefont
  {Profeta}, \citenamefont {Eremets},\ and\ \citenamefont
  {Goedecker}}]{Elemental_P-temperature_2017}%
  \BibitemOpen
  \bibfield  {author} {\bibinfo {author} {\bibfnamefont {J.~A.}\ \bibnamefont
  {Flores-Livas}}, \bibinfo {author} {\bibfnamefont {A.}~\bibnamefont {Sanna}},
  \bibinfo {author} {\bibfnamefont {A.~P.}\ \bibnamefont {Drozdov}}, \bibinfo
  {author} {\bibfnamefont {L.}~\bibnamefont {Boeri}}, \bibinfo {author}
  {\bibfnamefont {G.}~\bibnamefont {Profeta}}, \bibinfo {author} {\bibfnamefont
  {M.}~\bibnamefont {Eremets}}, \ and\ \bibinfo {author} {\bibfnamefont
  {S.}~\bibnamefont {Goedecker}},\ }\href@noop {} {\bibfield  {journal}
  {\bibinfo  {journal} {Physical Review Materials}\ }\textbf {\bibinfo {volume}
  {1}},\ \bibinfo {pages} {024802} (\bibinfo {year}
  {2017}{\natexlab{b}})}\BibitemShut {NoStop}%
\bibitem [{\citenamefont {Filip}\ \emph {et~al.}(2014)\citenamefont {Filip},
  \citenamefont {Eperon}, \citenamefont {Snaith},\ and\ \citenamefont
  {Giustino}}]{steric_gaps_NatCom2014}%
  \BibitemOpen
  \bibfield  {author} {\bibinfo {author} {\bibfnamefont {M.~R.}\ \bibnamefont
  {Filip}}, \bibinfo {author} {\bibfnamefont {G.~E.}\ \bibnamefont {Eperon}},
  \bibinfo {author} {\bibfnamefont {H.~J.}\ \bibnamefont {Snaith}}, \ and\
  \bibinfo {author} {\bibfnamefont {F.}~\bibnamefont {Giustino}},\ }\href@noop
  {} {\bibfield  {journal} {\bibinfo  {journal} {Nature communications}\
  }\textbf {\bibinfo {volume} {5}},\ \bibinfo {pages} {5757} (\bibinfo {year}
  {2014})}\BibitemShut {NoStop}%
\bibitem [{\citenamefont {Meloni}\ \emph {et~al.}(2016)\citenamefont {Meloni},
  \citenamefont {Palermo}, \citenamefont {Ashari-Astani}, \citenamefont
  {Gr{\"a}tzel},\ and\ \citenamefont {Rothlisberger}}]{meloni2016valence}%
  \BibitemOpen
  \bibfield  {author} {\bibinfo {author} {\bibfnamefont {S.}~\bibnamefont
  {Meloni}}, \bibinfo {author} {\bibfnamefont {G.}~\bibnamefont {Palermo}},
  \bibinfo {author} {\bibfnamefont {N.}~\bibnamefont {Ashari-Astani}}, \bibinfo
  {author} {\bibfnamefont {M.}~\bibnamefont {Gr{\"a}tzel}}, \ and\ \bibinfo
  {author} {\bibfnamefont {U.}~\bibnamefont {Rothlisberger}},\ }\href@noop {}
  {\bibfield  {journal} {\bibinfo  {journal} {Journal of Materials Chemistry
  A}\ }\textbf {\bibinfo {volume} {4}},\ \bibinfo {pages} {15997} (\bibinfo
  {year} {2016})}\BibitemShut {NoStop}%
\bibitem [{\citenamefont {Kresse}\ and\ \citenamefont
  {Furthm\"{u}ller}(1996)}]{VASP_Kresse}%
  \BibitemOpen
  \bibfield  {author} {\bibinfo {author} {\bibfnamefont {G.}~\bibnamefont
  {Kresse}}\ and\ \bibinfo {author} {\bibfnamefont {J.}~\bibnamefont
  {Furthm\"{u}ller}},\ }\href@noop {} {\bibfield  {journal} {\bibinfo
  {journal} {Comput. Mat. Sci.}\ }\textbf {\bibinfo {volume} {6}},\ \bibinfo
  {pages} {15} (\bibinfo {year} {1996})}\BibitemShut {NoStop}%
\end{thebibliography}%

\end{document}